\pdfoutput=1

\newcommand{\caproman}[1]{\uppercase\expandafter{\romannumeral#1}}

\newcommand{\BGmail}{bernardgottschalk\,@\,gmail.com}

\newcommand{\sect}[1]{Sec.\,\ref{sec:#1}}
\newcommand{\eqn}[1]{Eq.\,(\ref{eqn:#1})}

\newcommand{\fig}[1]{Fig.\,\ref{fig:#1}}
\newcommand{\tbl}[1]{Table\,\ref{tbl:#1}}

\documentclass[12pt]{article}

\title{\bf Rapid measurement of tension\\in multi-wire arrays\\using free damped oscillations}
\author{B. Gottschalk\thanks{Harvard University Laboratory for Particle Physics and Cosmology (LPPC), 18 Hammond St., Cambridge, MA 02138, USA, \BGmail}}
\date{\today}

\usepackage{amsmath}
\usepackage{latexsym}
\usepackage{graphicx}
\usepackage{rotating}
\usepackage{lineno}
\usepackage{setspace}
\usepackage[linktocpage=true,pdftex,hyperfigures=true]{hyperref}

\begin{document}

\maketitle
\vspace{.15in}
\begin{center}{\large\bf for Alan Cromer (1935$-$2005)}\end{center}

\vspace{.30in}
\begin{center}{\em Did you try whacking it?} --- R.V. Pound\end{center}

%\vspace{1in}\begin{center}{\Large\bf DRAFT}\end{center}
%\enlargethispage*{1000pt}

\clearpage
\begin{abstract}
\noindent We explore the possibility of using free (as opposed to driven) oscillations to verify wire tension in large, open-frame multiwire planes.

Using finite-element simulation we predict the signal when two wires are set in motion by applying and holding a DC voltage between them ({\em electrostatic excitation}). The resulting current signal proves to be marginal, for a realistic voltage, even for the longest wires foreseen, and out of reach for the shortest. However, the program finds the relation between voltage, oscillation amplitude and signal, and it is easy to get enough amplitude (still $\ll1$\,mm) by {\em mechanical excitation} (gently tapping the frame).

We perform an end-to-end experimental test. Four 1.5\,m wires, stretched to various tensions around 5\,N, are mounted in an aluminum channel. They are connected to a low-noise, balanced current-to-voltage preamplifier designed for good common-mode rejection. The 150\,V bias, from 9\,V batteries in series, is built into the preamp.

Data are taken with a Tektronix 1202B scope, stored on a USB flash drive, and transferred to a laptop computer for analysis. (Eventually, the scope would be programmed as an extension of the computer.) Ten measurements, each comprising 2500 samples spanning 1\,s, are taken for each of the three wire pairs. The wires are excited by tapping the frame with a small mallet. Because the wire tensions are different, the waveforms exhibit beats.

The data are analyzed by a program which, rather than FFT or nonlinear least-squares, simply decomposes the waveform step by step to extract the base and beat frequencies and thereby, two frequencies per wire pair. Finally, those frequencies are sorted and averaged as appropriate to disambiguate the resonant frequency for each wire, which is then converted to a tension.

To close the loop, the four tensions are found mechanically by measuring the g/mm required to deflect each wire at its midpoint. The two values of tension agree well for the 1\,s measurements. We conclude that the method verifies tension to approximately $\pm2\%$ in a little over 1\,s for 1.5\,m wires, and faster for shorter wires.

The real-world performance vis-a-vis driven oscillations awaits more realistic tests, and will also depend on the workflow and objectives in fabricating, testing and installing the wire frames.  

 \end{abstract}

\clearpage\tableofcontents

\clearpage
\section{Introduction}\label{sec:Introduction}
Modern-day physics experiments sometimes use detectors with tens of  thousands of stretched wires. For quality assurance, one needs to measure the wire tension. Direct measurement of the deflection produced by a known force at (say) mid-wire is time consuming, and may be impossible if there are interfering wire layers. Instead, one measures the fundamental frequency of mechanical vibration namely
\begin{equation}f\;=\;\frac{1}{2L}\sqrt{\frac{T}{\rho}}\qquad\hbox{Hz}\label{eqn:f}\end{equation}
where $L$ (m) is the length of the wire, $T$ (N) is the tension and $\rho$ (kg/m) is the linear density of the wire.

$f$ may be measured either by searching for a resonance in driven oscillations or by studying the free damped oscillations following an initial perturbation. The literature (which is extensive) seems to favor the former, even though the latter is potentially much faster. The method to be described will take just over one second to acquire and analyze the data per pair of 1.5\,m wires. 

The frequency-search technique is exemplified by recent work of Garcia-Gamez et al. \cite{Garcia-Gamez2019}. Two wires flanking the wire under test are subjected to a combination of DC and AC voltages, and the frequency of the AC is swept to detect the resonance. The signal is derived from the change of capacitance of the wire under test to the other two, as in a capacitance microphone.

The other approach is described (for instance) by Lang et al. \cite{Lang1999}. They place the wire in a magnetic field, excite it using a current pulse, and subsequently measure the signal generated by the wire moving in the magnetic field. The signal is small, and the technique takes advantage of the electrostatic shielding of the conducting `straw' in which the wire is contained. To cite another example, Brinkley et al. \cite{Brinkley1996} measure damped oscillations, excited by pulsing the field-shaping electrodes, of single wires in conducting drift tubes.

We are concerned, instead, with measuring tension in large open-frame wire planes such as those envisioned for the DUNE liquid-argon detector. Applying a sufficient magnetic field would be difficult, electrostatic shielding is non-existent, and applied voltage must respect the spacing between solder pads.

We propose simply to apply and hold a DC voltage between two adjacent wires, with all the others floating. That excites damped harmonic oscillations, with the wires eventually reaching a static equilibrium position where they are slightly closer together. We call that {\em electrostatic excitation}. During the oscillations their mutual capacitance oscillates slightly. Because the voltage is fixed, that results in a small AC current that can be measured by suitable external circuitry.

If the tension in the two wires is slightly different, beats ensue, and two resonant frequencies can be deduced from the waveform. We do not know which frequency corresponds to which wire. However, when we advance by one wire, we will find (hopefully) that one of those frequencies matches one found previously, resolving the ambiguity. Moving through the entire plane, we will have measured each wire twice, except the first and last wires, which are measured once.
%\enlargethispage{2\baselineskip}

In this report we describe an end-to-end demonstration of this strategy, with one major shortfall. Initially we hoped to use electrostatic excitation, as described above. With the DC bias voltage and preamp design we have used so far, the resulting signal-to-noise ratio is too small. For wires shorter than the 1.5\,m tested, our paper study shows that electrostatic excitation is even further out of reach. However, we found that suitable vibrations could be excited reliably by simply tapping the side of the frame ({\em mechanical excitation}). The resulting oscillations, visible only as a faint blurring, are still extremely small. 

Since that may well be a path forward for any wire length, we have chosen to write up what we have done so far. \sect{simulation} describes a finite-element computation to estimate the signal frequency and amplitude and their variation with wire length, tension, applied voltage and other parameters. \sect{Hardware} describes the hardware: a four-wire test jig, with tensions deliberately varied to provide a more stringent test, and a purpose-built balanced preamplifier. (Because each wire acts as an antenna, single-ended amplification, with one wire biased and the other read out, is hopeless.) \sect{procedure} describes the measurement procedure and \sect{analysis} describes a somewhat lengthy, but nevertheless fast, analysis procedure. 

There is no serious doubt that $f$ is related to wire tension via \eqn{f}. Nevertheless, in our setup it was fairly easy to measure $T$ directly. We describe that in \sect{direct}, closing the loop. \sect{discussion} is a summary and discussion.

\section{Simulation}\label{sec:simulation}
\begin{figure}[p]
\centering\includegraphics[height=3.5in]{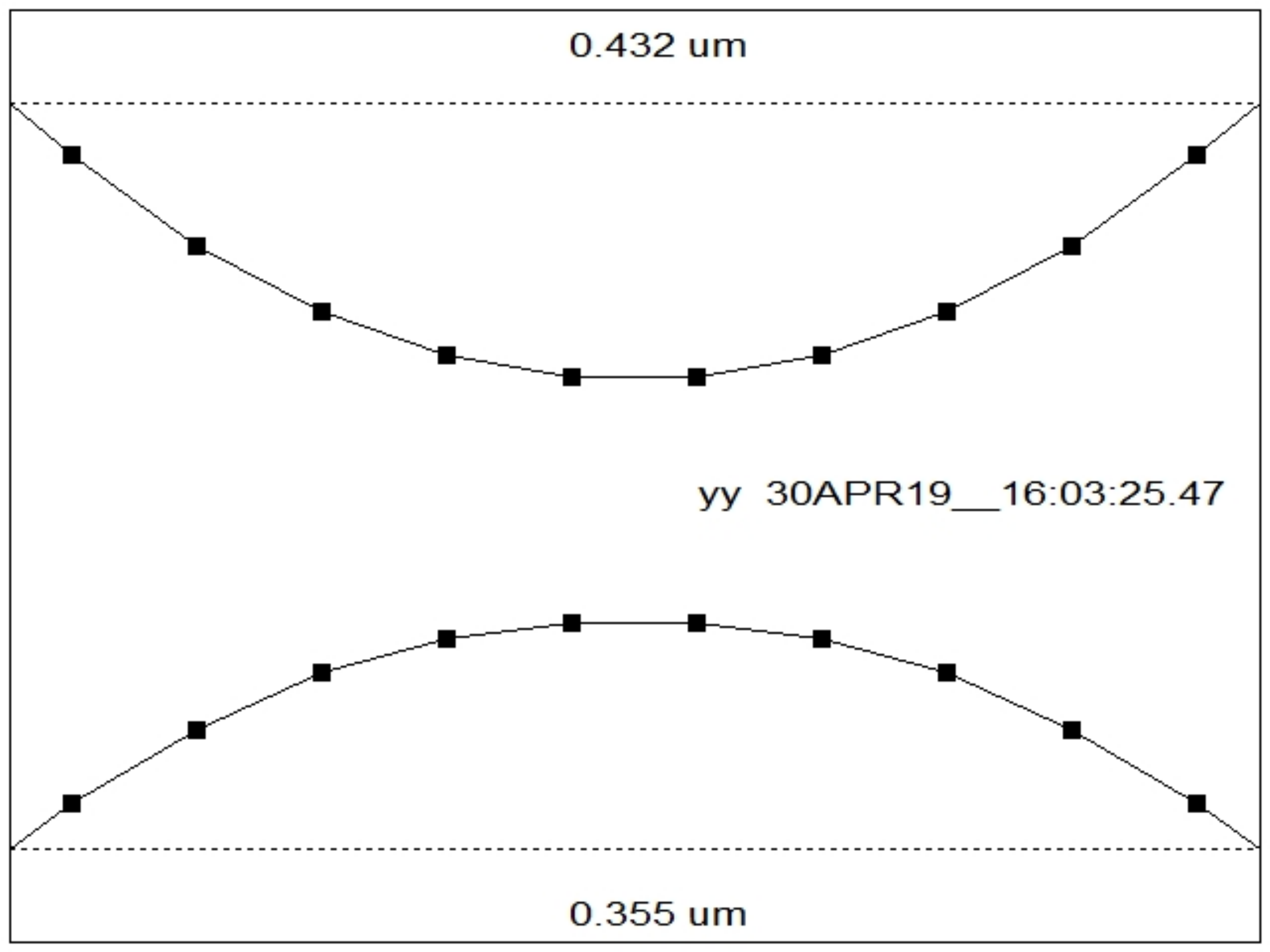}
\caption{Wire shapes during simulation. The transverse scale is greatly distorted: the wires are 5\,mm apart whereas the vibration amplitude is measured in $\mu$m.}\label{fig:ShowFrame}
\end{figure}

Rather than attempt a solution in closed form we wrote a finite-element program, one branch of a general Fortran program \texttt{WIRES} comprising all the routines required by this project. \fig{ShowFrame} illustrates the model: $N$ equal point masses $m_i$ for each wire, connected by massless conducting wires of finite diameter $d$. Each mass commands a length $\Delta x=L/N$ centered on that mass. Transverse mass displacements $y_i$ are measured from the rest position of each wire, positive inwards. 

\subsection{Formulas}\label{sec:Formulas}
The effective (that is, mean) transverse wire position corresponding to $m_i$ is
\begin{equation}\label{eqn:yeff}
y_\mathrm{\,eff}=0.125\;(y_a+6y_i+y_b)\quad,\quad1<i<N
\end{equation}
where
\begin{equation}\label{eqn:ya}
y_a=-y_1\quad(i=1),\quad=y_{i-1}\quad(i>1)
\end{equation}
and
\begin{equation}\label{eqn:yb}
y_b=-y_N\quad(i=N),\quad=y_{i+1}\quad(i<N)
\end{equation} 
The effective distance $D_i$ between the wire segments corresponding to each opposite mass is the wire pitch less the two values of $y_\mathrm{eff}$. The electrostatic force between those segments, if $\lambda_1,\lambda_2$ are linear charge densities (equal in this case), is
\begin{equation}\label{eqn:Fi}
F_i\;=\;\frac{1}{4\pi\epsilon_0}\frac{2\lambda_1\lambda_2}{D_i}\,\Delta x
\end{equation}
The mutual capacitance of the segments is
\begin{equation}\label{eqn:Ci}
C_i\;=\;\frac{\pi\epsilon_0}{\ln\,(2D_i/d)}\,\Delta x\quad,\quad D_i\gg d
\end{equation}
and the (equal and opposite) charge on each wire segment is
\begin{equation}\label{eqn:Qi}
Q_i\;=\;C_i\,V
\end{equation}
As $C_i$ oscillates so must $Q_i$ and therefore a current must flow out of one wire and into the other. The total charge is obtained by summing over $i$ and the current, by differentiating with respect to time (that is, simply dividing the total charge by the time per step).

Eqs. (\ref{eqn:yeff}) through (\ref{eqn:Qi}) allow us to find the (attractive) electrostatic force $F_{i\,\mathrm{,\,el}}$ on each mass given the positions of all the masses. The (repulsive) restoring force due to tension, in small-angle approximation, is
\begin{equation}
 F_{i\,\mathrm{,\,res}}\;=\;\frac{T}{\Delta x}\,(2\,y_i-ya-yb)
\end{equation}
and we assume an arbitrary damping force of the form
\begin{equation}
F_{i\,\mathrm{,\,damp}}\;=\;c_\mathrm{damp}\;v_i
\end{equation}
where $v_i$ is the velocity of $m_i$ and $c_\mathrm{damp}$ is an input parameter given in $\mu$N/(m/s). The total force on $m_i$ is
\begin{equation}
F_i\;=\;F_{i\,\mathrm{,\,el}}-F_{i\,\mathrm{,\,res}}-F_{i\,\mathrm{,\,damp}}
\end{equation}
The foregoing reasoning is heuristic since Eqs. (\ref{eqn:Fi}) and (\ref{eqn:Ci}) apply only to infinite wires. We have in essence assumed that end effects vanish for abutting segments and small displacements.

\subsection{Integration}
\begin{figure}[p]
\centering\includegraphics[height=3.5in]{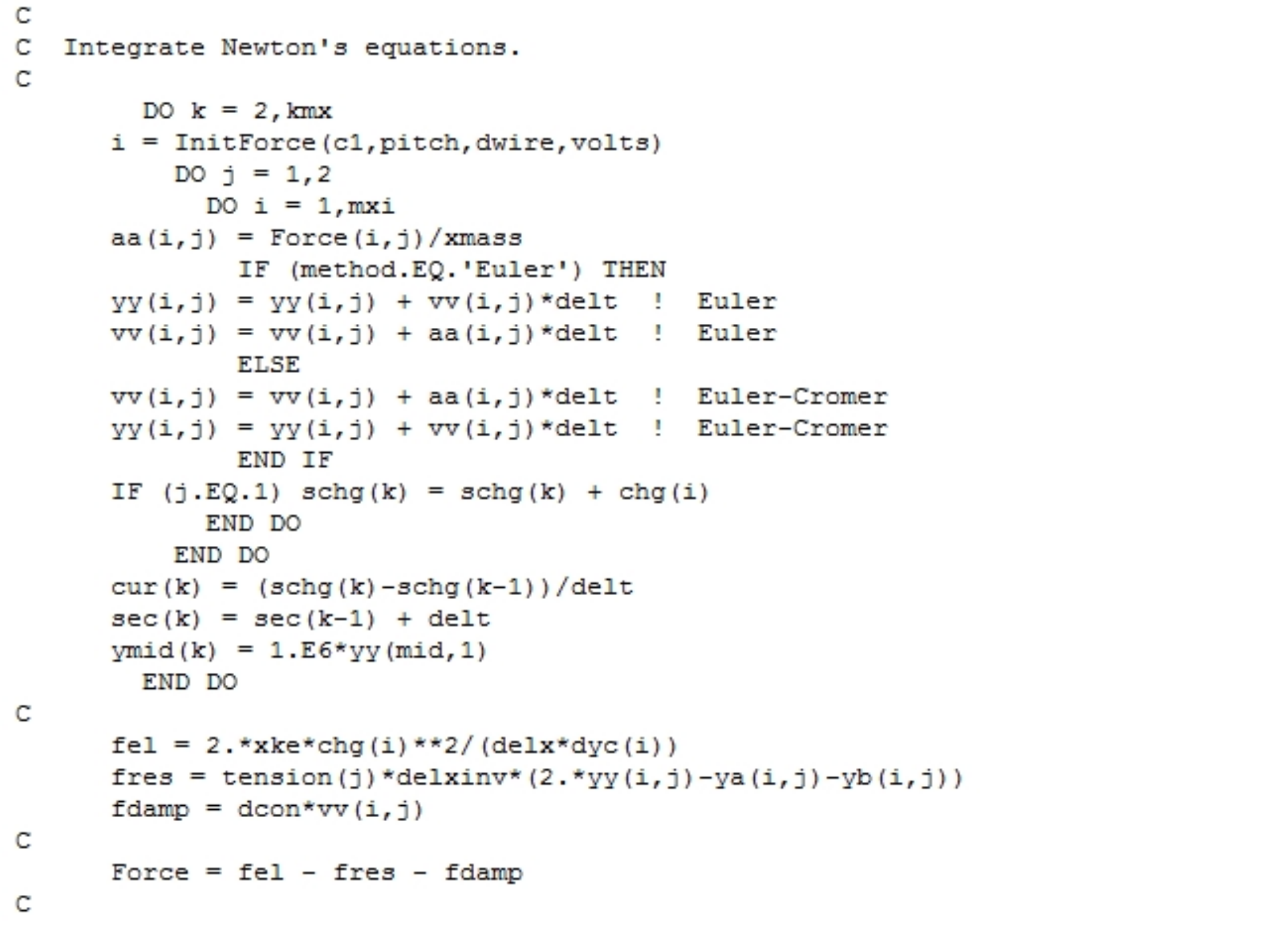}
\caption{Program fragment contrasting the Euler and Euler-Cromer approximations. Two lines are simply swapped.}\label{fig:Fragment}
\end{figure}

After initializing positions, velocities and accelerations we need to integrate Newton's equations of motion for each $m_i$. There are many algorithms, some quite complicated. Fortunately, in this case the amazingly simple Euler-Cromer algorithm mentioned almost in passing by Gould et al. \cite{Gould2005} appears to work well. See \fig{Fragment} which illustrates the difference between it and the Euler algorithm (which diverges after a few cycles). We simply swap two lines. In a brief and entertaining paper, Cromer \cite{Cromer1981} attributes its accidental discovery to a high-school student and explains why it works so well for periodic systems.

The algorithm fails if the number of steps per cycle is less than 30 or so. For that reason the input explicitly specifies the number of cycles desired and the steps per cycle. Other parameters needed are worked out from those using \eqn{f}. Ten masses per wire is enough, and is built into the program (\texttt{mxi}). If that is significantly increased by recompiling the program, steps/cycle needs to be increased as well. 

\subsection{Example}
\begin{figure}[p]
\centering\includegraphics[height=3.5in]{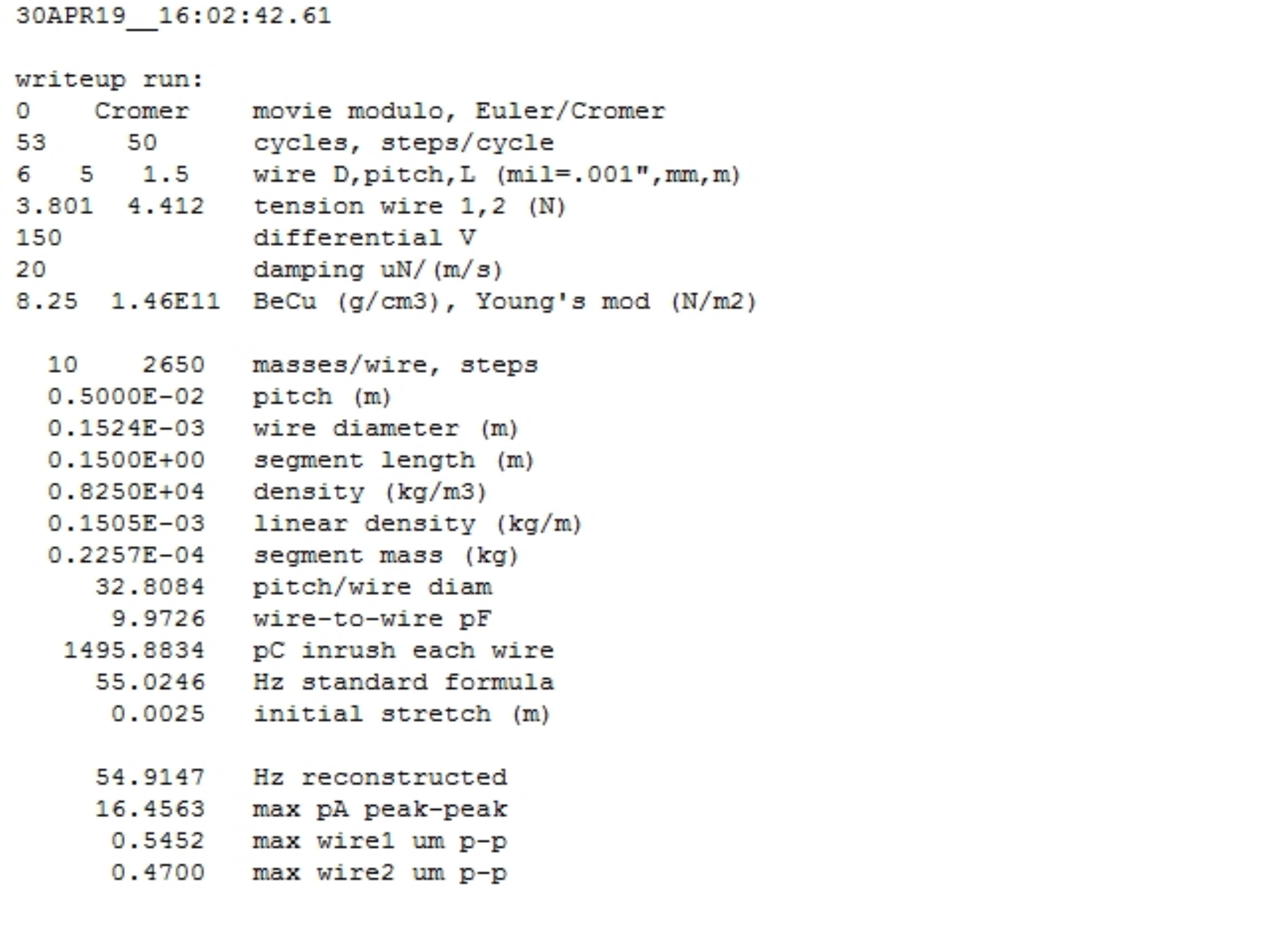}
\caption{Numerical input and output from simulation program.}\label{fig:WiresOUT}
\end{figure}

As an example, we simulate one of the configurations measured later (gap 3). The three blocks of \fig{WiresOUT} show the input (itself a text file), intermediate output and simulator output. Note the good agreement between simulated vibration frequency and the standard formula \eqn{f}. Note also that, with 150\,VDC applied wire-to-wire, the peak-to-peak (p-p)  vibration amplitude is of order 0.5\,$\mu$m or 0.01\% of the pitch, and the corresponding p-p current is only 16\,pA.

\begin{figure}[p]
\centering\includegraphics[height=3.5in]{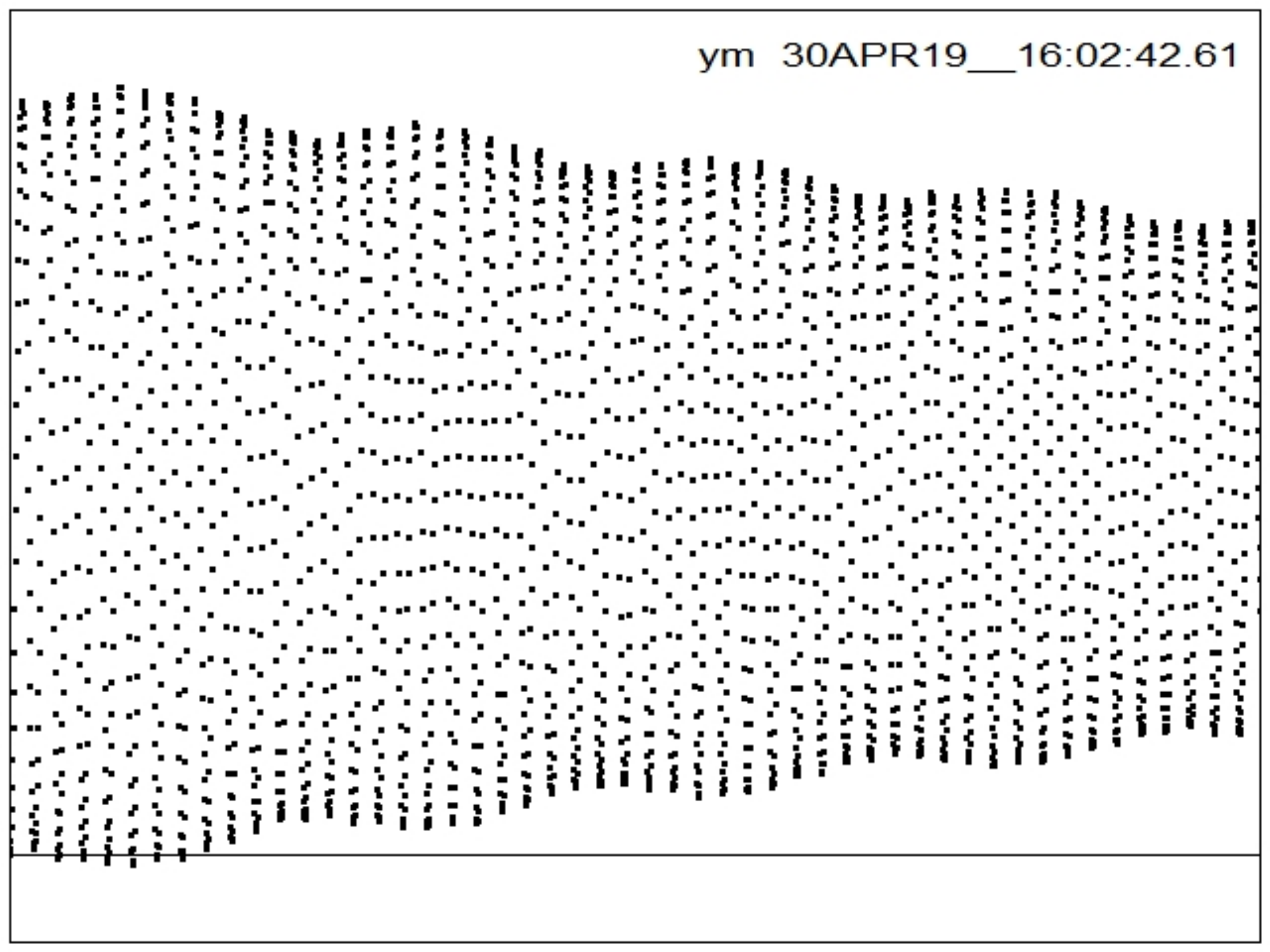}
\caption{Displacement of midpoint, wire 1. The slight oscillations in the envelope are due to the bounded error in the Euler-Cromer algorithm, not beats.}\label{fig:ym}
\end{figure}

\begin{figure}[p]
\centering\includegraphics[height=3.5in]{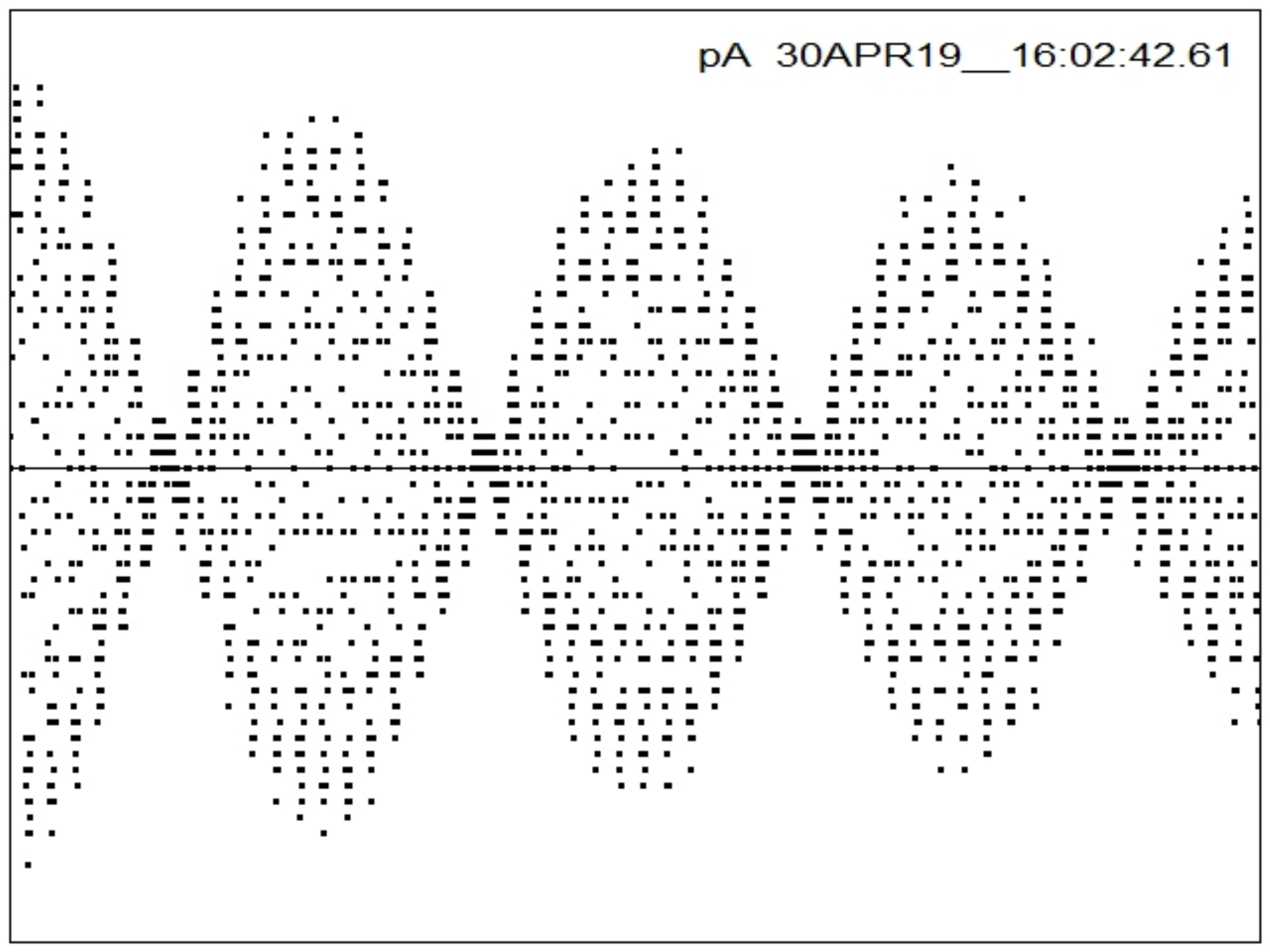}
\caption{Differential current at fixed voltage, simulated run\,67.}\label{fig:pA}
\end{figure}

\begin{figure}[p]
\centering\includegraphics[height=3.5in]{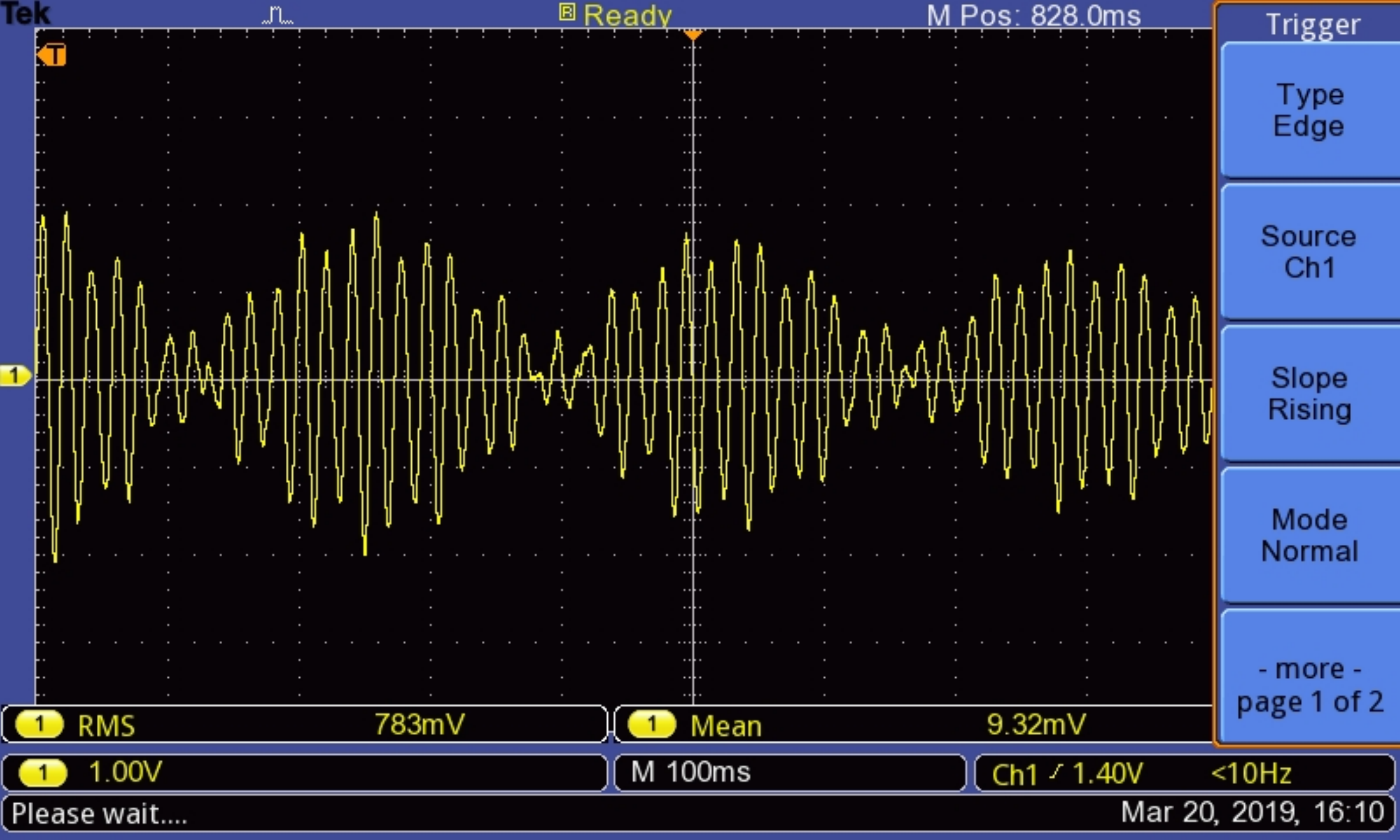}
\caption{Scope trace for simulated run 67. (Trigger menu obscures part of the trace.)}\label{fig:F0067TEK}
\end{figure}

\fig{ym} shows the oscillations of the midpoint of wire\,1 about the equilibrium value. The horizontal line is the resting state of the wire at $V=0$. The slight waves in the envelope reflect the bounded error in the Euler-Cromer algorithm (not beats). \fig{pA} shows the differential signal. \fig{F0067TEK} is the oscilloscope signal in the corresponding measurement. Considering that the true parameters are subject to experimental error, and that part of the oscilloscope trace is covered by the trigger menu, the agreement is not bad. 

\subsection{Scaling}
\begin{figure}[p]
\centering\includegraphics[height=3.5in]{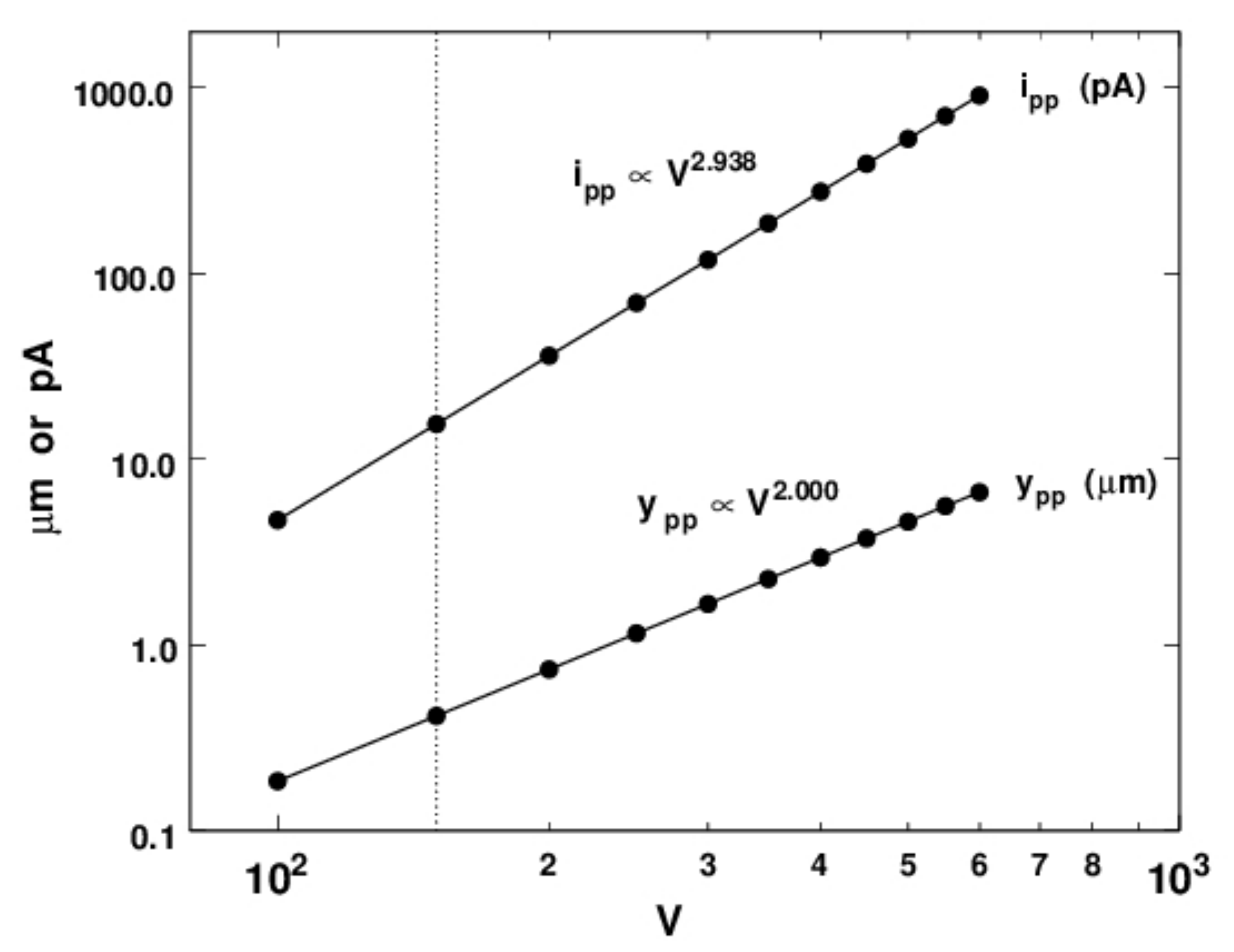}
\caption{Scaling of peak-to-peak displacement and current with voltage. The vertical line marks the voltage used in the experiment.}\label{fig:Vscaling}
\end{figure}

The variation of signal with $V$ is of interest because that is one of the parameters we can control, within reason. \fig{Vscaling} shows that the amplitude of vibration (assuming electrostatic excitation) is proportional to $V^2$ and the signal to $V^{2.938}$. This (except for the slight departure from $V^3$) can be understood using a simple analytical model.

Consider a parallel plate capacitor of area $A$ with the upper plate fixed at $y=D$ and grounded and the lower (initially at $y=0$) attached to a spring of spring constant $k$. Gravity is switched off, or the plates are massless. A voltage $V$ is applied to the lower plate which immediately (before the plate moves appreciably) acquires a charge $Q_1$. The plate is attracted upwards and eventually settles down at $y\ll D$ where the upwards electrostatic force equals the spring force $ky$ and the charge is $Q_2$. (Imagine the system to be heavily damped, or $y$ to be the equilibrium position after very many oscillations.) The initial capacitance of the parallel-plate capacitor is \cite{Purcell}
\begin{equation}\label{eqn:PPC}
C_1=\frac{\epsilon_0 A}{D}
\end{equation}
and the electrostatic force on either plate immediately afterwards is
\begin{equation}\label{eqn:force}
F=\frac{\epsilon_0 A V^2}{2D^2}
\end{equation}
At equilibrium
\[ky=\frac{\epsilon_0 A V^2}{2(D-y)^2}\]
or
\begin{equation}\label{eqn:y}
y\;=\;\frac{\epsilon_0 A}{D}\frac{V^2}{2kD}\;=\;\frac{C_1}{2kD}\;V^2\quad (y\ll D)
\end{equation}
The second form shows that the derivation applies to any rigid capacitor (e.g. parallel rods as considered in \sect{Formulas}). To obtain the scaling for charge or current, consider
\begin{eqnarray}
Q_2-Q_1&=&(C_2-C_1)V\nonumber\\
&=&(\frac{\epsilon_0 A}{D-y}-\frac{\epsilon_0 A}{D})\;V\nonumber\\
&=&\frac{\epsilon_0 A}{D^2}\;y\;V\quad (y\ll D)\nonumber\\
\noalign{\noindent and from \eqn{y}}
Q_2-Q_1&=&\frac{C_1^2}{2kD^2}\;V^3\quad (y\ll D)
\end{eqnarray}
The slight departure from $V^3$ in the simulation is not due to the difference between parallel plate and parallel rod geometry, but instead to the fact that the wires are not rigid rods (transverse displacement depends on $x$).

\begin{figure}[p]
\centering\includegraphics[height=3.5in]{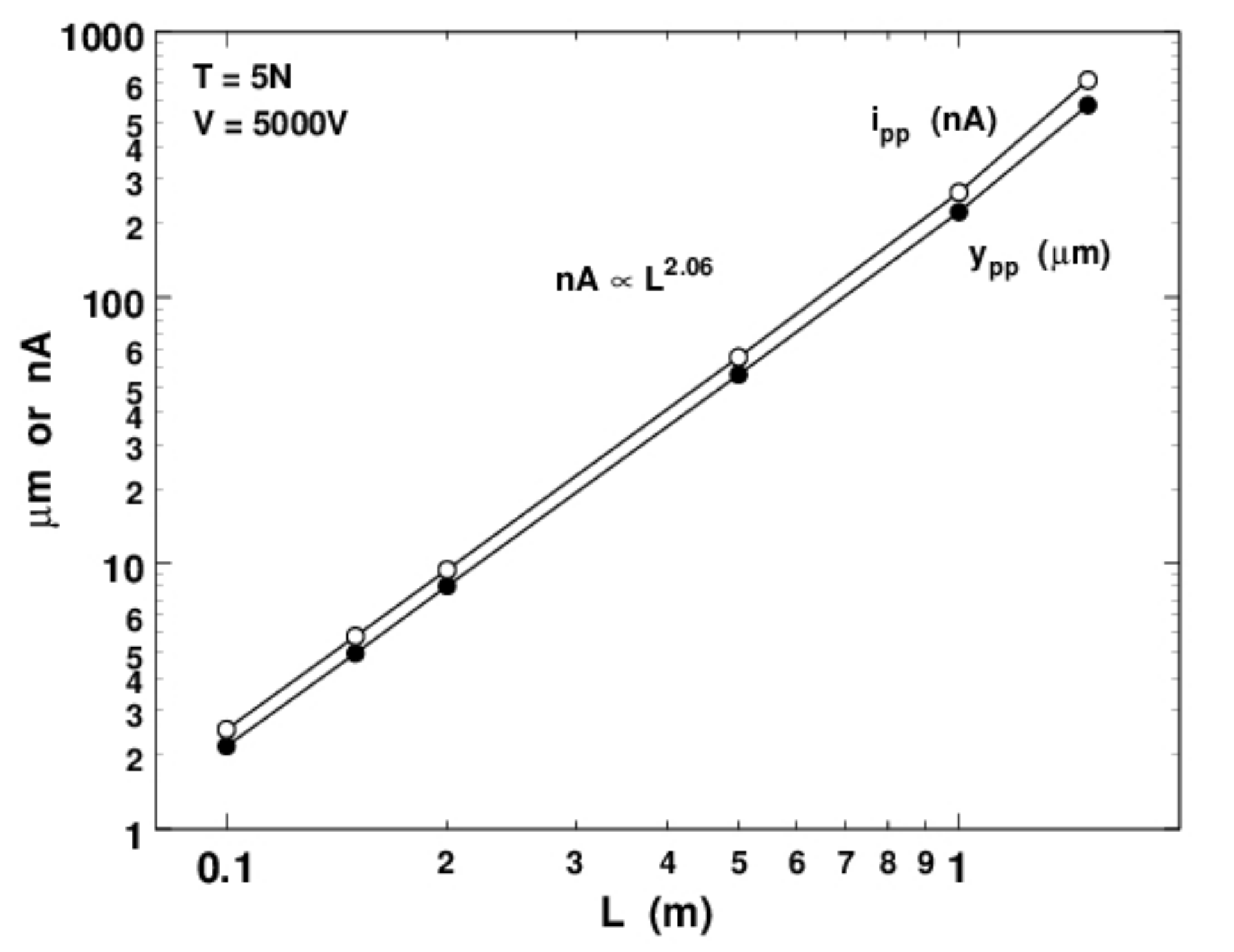}
\caption{Scaling of peak-to-peak displacement and current with wire length.}\label{fig:Lscaling}
\end{figure}

\fig{Lscaling} shows the scaling with wire length. To get a reasonable signal for short wires and electrostatic excitation we have used an unrealistic 5000\,V. The message from this plot is that electrostatic excitation is not feasible for short wires.

The good news, however, is this. If, using electrostatic excitation at 150\,V, we obtain 16\,pA p-p at a vibration amplitude 0.5\,$\mu$m p-p (cf. \fig{WiresOUT}) then, if we attain 50\,$\mu$m by mechanical excitation we will see 1600\,pA, easily measured, at the same voltage. And that is still only 0.05\,mm or 1\% of the pitch, barely visible.

\section{Hardware}\label{sec:Hardware}

\subsection{Test Jig}
\begin{figure}[p]
\centering\includegraphics[height=3.5in]{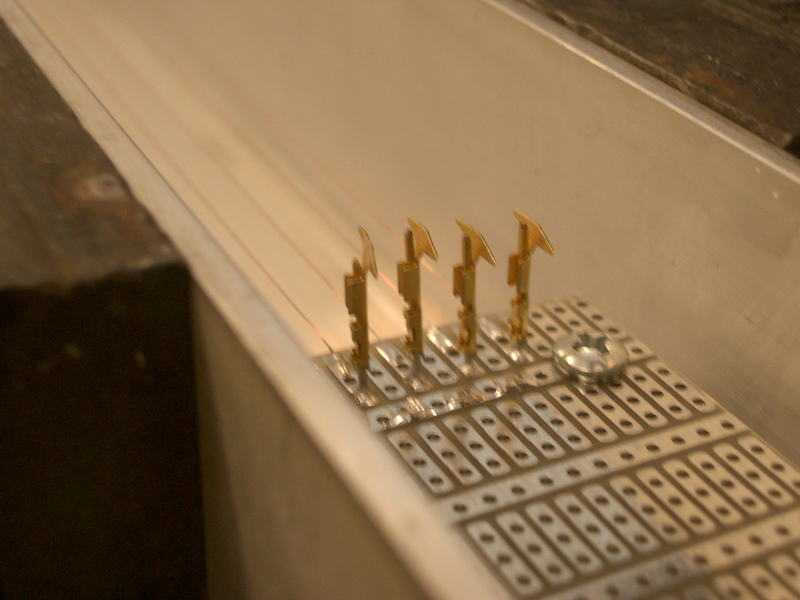}
\caption{Solder pads, showing wires and square pins for probe attachment.}\label{fig:pads}
\end{figure}

We stretched four $0.006\,''$ diameter 1.5\,m BeCu wires at 5\,mm pitch inside a length of $2\,''\times2\,''$ aluminum channel. The tension was about 1\,pound (deliberately varied somewhat) measured with a fisherman's scale. The wires were soldered to the pads of circuit board scraps, with connecting pins at one end of the wire (\fig{pads}). This arrangement protected the wires and was easy to stow and deploy. It provides more shielding than the real thing, but there was still plenty of electromagnetic interference (EMI), especially 60\,Hz, as shown by detuning the balance adjustment discussed later.

\subsection{Preamplifier}
\begin{figure}[p]
\centering\includegraphics[height=3.5in]{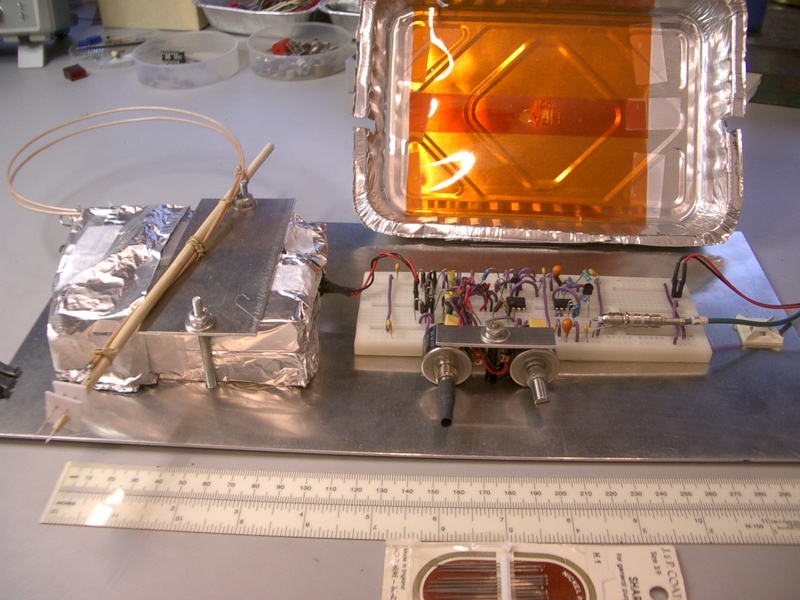}
\caption{Open preamp. From left: probe, Teflon cable, wrapped and grounded 75\,V battery packs, preamp on breadboard with BALANCE and OFFSET adjustments.}\label{fig:openPreamp}
\end{figure}

\begin{figure}[p]
\centering\includegraphics[height=3.5in]{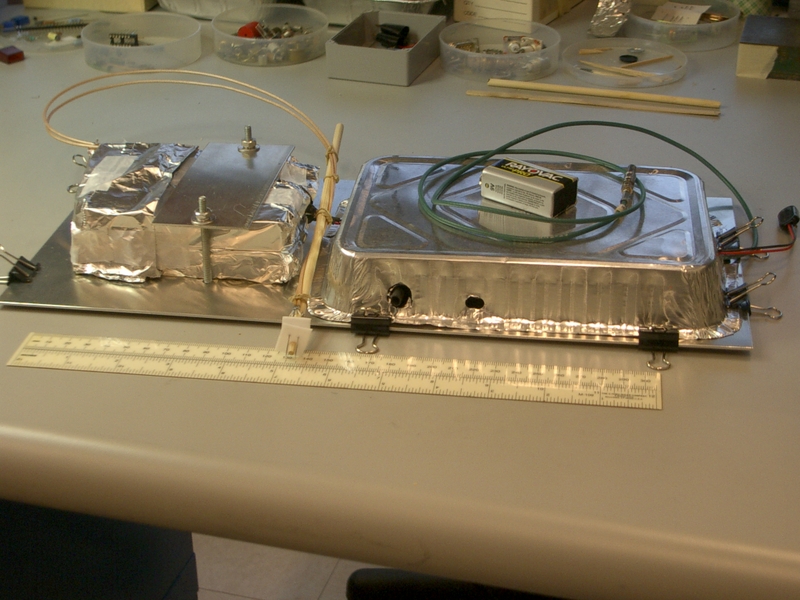}
\caption{Shielded preamp.}\label{fig:closed2}
\end{figure}

\begin{figure}[p]
\centering\includegraphics[height=3.5in]{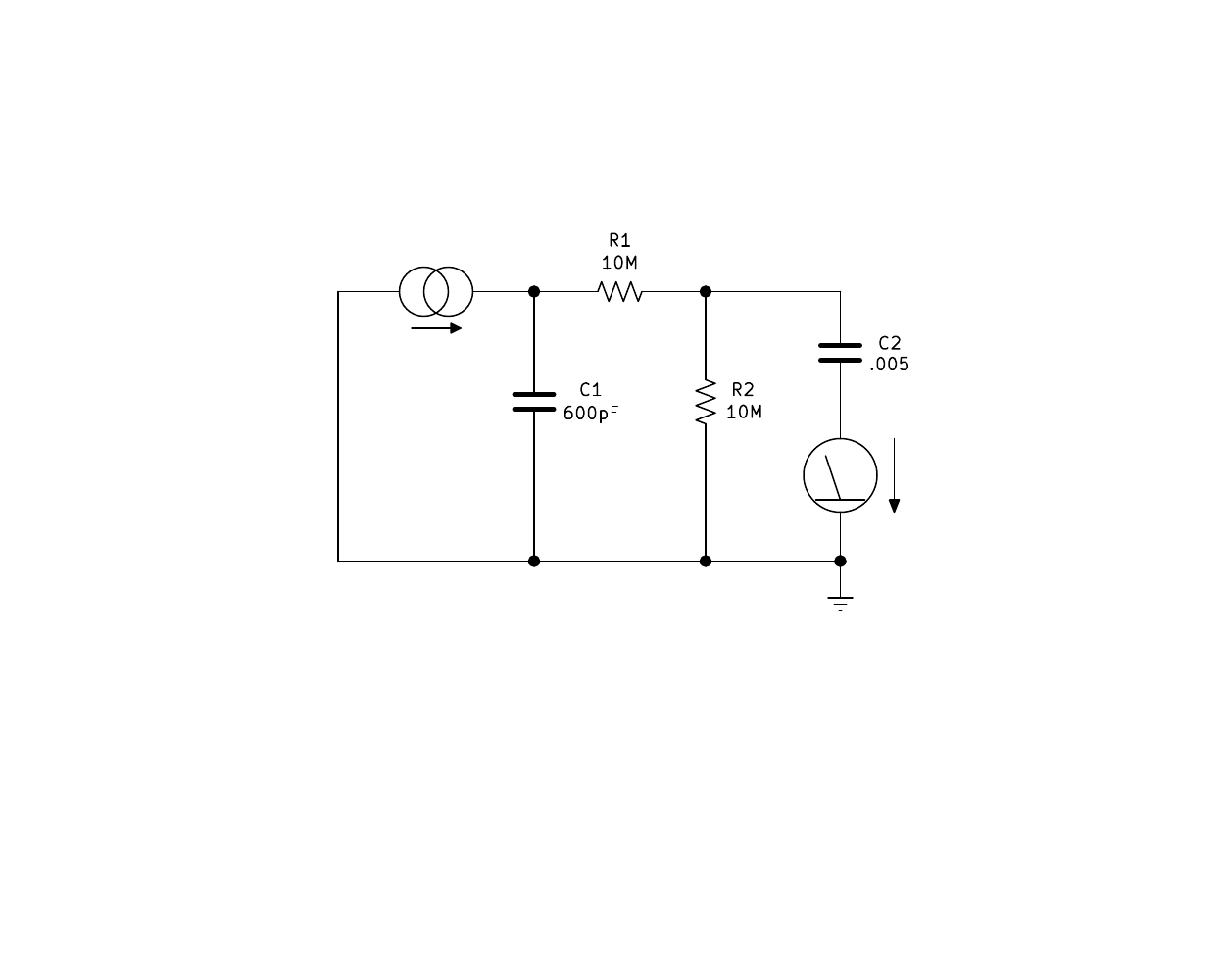}
\caption{Equivalent circuit, preamp front end.}\label{fig:equivCircuit}
\end{figure}

\begin{sidewaysfigure}[p]
\centering\includegraphics[width=9.5in]{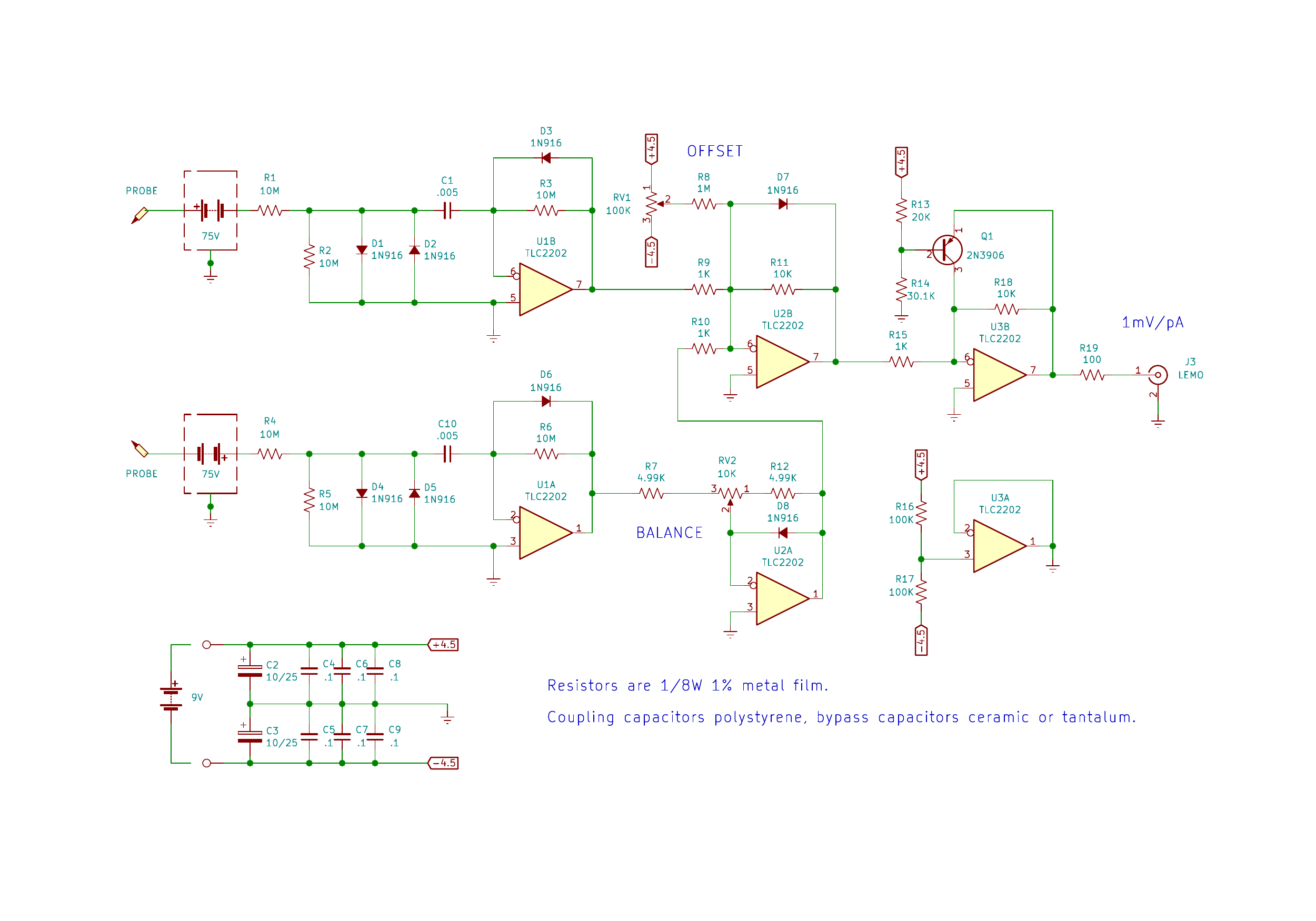}
\caption{Differential current-to-voltage amplifier with battery packs.}\label{fig:WireAmp}
\end{sidewaysfigure}

We used a legacy breadboard and through-hole components (Figs. \ref{fig:openPreamp} and \ref{fig:closed2}). \fig{equivCircuit} is the equivalent circuit of the front end, discussed later, and \fig{WireAmp} is a complete schematic. The circuit shown and described here is that used for the 1\,s measurements presented later, without the modifications we might make based on what we learned. In particular, the preamp was originally designed for end-to-end DC coupling, which explains the unconventional bias arrangement. However, with 150\,V applied between adjacent pads, unstable parasitic currents (even after cleaning) rule out DC coupling. Once coupling capacitors (C1 and C10, \fig{WireAmp}) are introduced, a much more straightforward biasing scheme would be possible.

To avoid noise from power supplies, both preamp and 150\,V bias were supplied by 9\,V batteries. That is perfectly practical: the bias cells should last for their shelf life (years) and an alkaline cell powering the preamp, for about 30\,hours.

Returning to \fig{WireAmp}, the main design problems are noise, common-mode response (the common-mode signal from EMI being many orders of magnitude larger than the differential current signal), and rapid recovery from the inrush signal when bias is first applied.\footnote{~May be irrelevant for mechanical excitation.} 

Recovery is addressed by keeping each opamp within its dynamic range, using diodes in the feedback loop of early stages and the clamp on U3B later.\footnote{~With legacy glass-body diodes, any diode is a photodiode at the pA level! We covered them with black heat-shrink tubing.}

Common-mode rejection is improved by using an inverter and summing amplifier (U2A,B) instead of the conventional difference amplifier configuration, which relies on common-mode rejection by the opamp.\footnote{~All the opamps are used in inverting mode, where common-mode rejection is irrelevant.} Rather than matching components in the front end, we used a balance adjustment RV2.\footnote{~With this detuned, the output noise is mostly 60\,Hz as expected. At the optimal setting there is a very small residual 120\,Hz component, at least in our lab.} The input offset stability of the TLC2202 is sufficiently good that a single offset adjustment RV1 keeps the output DC voltage reasonably small.\footnote{~Rate dependent baseline shifts introduced by multiple coupling capacitors would be a more serious problem than a slowly varying DC level, which is easily taken care of by the analysis program.} U3A is a supply splitter, allowing operation from a single 9\,V battery.

\begin{figure}[p]
\centering\includegraphics[height=3.5in]{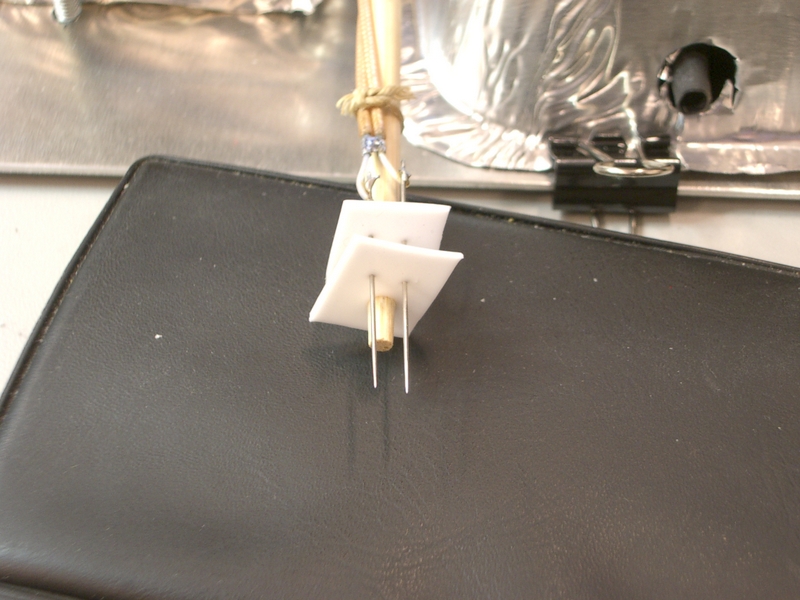}
\caption{Probe.}\label{fig:probe}
\end{figure}

As always, the front end is critical. (We repeat that we describe it {\em as used}, not as we might change it in the light of developments.) If the probe  is to be used with test leads, low triboelectric effect in both is essential to avoid noise whenever the probe is moved. Teflon is by far the best insulator in this regard, so the probe itself consists of sewing needles held by two Teflon spacers and connected to the preamp by individually shielded Teflon RG178B/U coaxial cables (\fig{probe}).

We found that each battery pack required a separate grounded shield to avoid pickup as well as parasitic currents between the batteries. Because of the size of the assembly, that results in a rather large capacitance to ground, about 0.6\,nF, whose effect we discuss now. \fig{equivCircuit} is the equivalent front-end circuit of one of the current-to-voltage (`transimpedance') stages e.g. U1B. Functionally, C1 represents the parasitic capacitance to ground of the shielded batteries; R1 limits the inrush current as well as the DC current if the probe is accidentally shorted; R2 completes the DC path to ground, and the coupling capacitor C2 should be chosen as described below. The current meter at the ground end of C2 serves only to emphasize that the (time varying) current through C2 is the `output' of this stage.\footnote{~The ground is virtual, assuming an ideal opamp. R3 (\fig{WireAmp}) serves merely to convert the current to a voltage, and is therefore not part of the equivalent circuit.} 

A qualitative discussion will be more useful than a full-bore analysis using complex impedances etc. At high frequency the signal current prefers C1 to R1. In other words the combination is a low-pass filter of time constant $\tau_1=R_1C_1\approx6$\,ms. Similarly, at high frequency the current prefers C2 to R2 so that, insofar as output is concerned, we have a high-pass filter of $\tau_2=R_2C_2\approx50$\,ms. The upshot is a band-pass filter of time constant $(\tau_1\tau_2)^{1/2}\approx17$\,ms which is near the period of the oscillations we plan to measure. For shorter wires and higher frequencies this would have to be improved. To avoid rate-dependent baseline shifts, C2 should be no larger than necessary for the lowest frequency encountered.
\enlargethispage{\baselineskip} 

\section{Procedure}\label{sec:procedure}
We acquired data using a Tektronix TBS1202B scope. Eventually one would program the scope to act as an extension of the computer via USB but, not wishing to be sidetracked into that project, we saved the data to a USB flash drive and transferred them manually to a Lenovo T400 laptop.

It proved necessary to connect the aluminum channel, preamp ground and scope ground. In the open-frame geometry foreseen, there is no shield and therefore nothing to ground, so that is a potential problem. We conjecture that a partial floating shield is worse than no shield if, as is likely, the coupling to EMI is electrostatic. The shield picks up far more than the wires, and transmits it to the wires in an unbalanced way. However, repeating the experiment on the real thing is the only definitive test.

Exploratory runs showed that electrostatic excitation yielded too little signal. However, both the preamp and the test wire jig proved extremely microphonic, and we soon found that tapping the jig at midpoint with a finger yielded good signals. Eventually, a small improvised mallet proved more reproducible. We recorded 30 runs using a sweep duration of 2.5\,s. The scope takes 2500 samples, whatever the sweep speed. The 2.5\,s runs were analyzed, but proved disappointing (see below).

\begin{figure}[p]
\centering\includegraphics[height=3.5in]{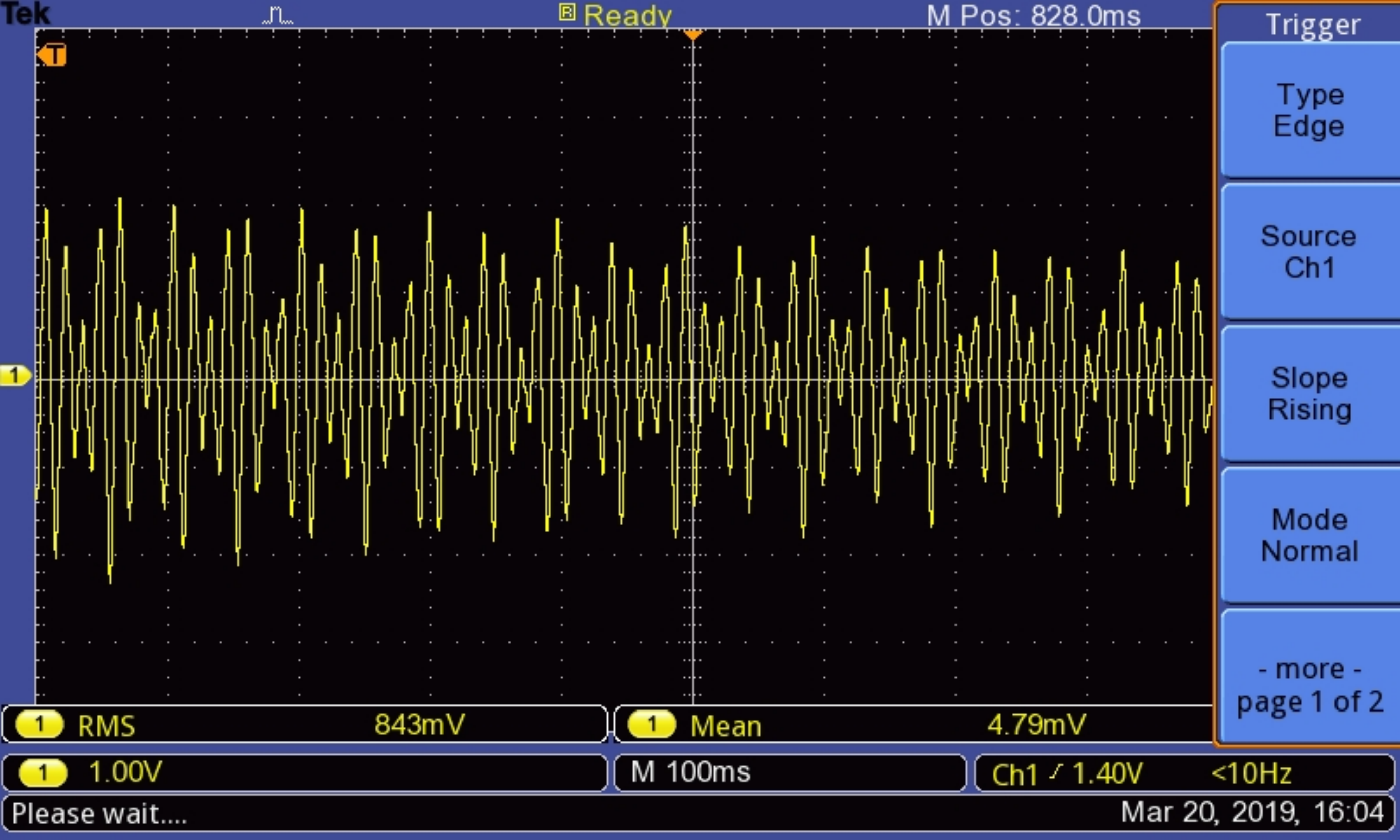}
\caption{Scope trace for run 60, our example.}\label{fig:F0060TEK}
\end{figure}

The final 30 runs, at 1\,s sweep duration, were obtained by setting the trigger fairly high (1.5\,V out of a possible 3.5\,V), tapping the jig with the mallet, ignoring the first sweep (which overloads the preamp) and saving the second. \fig{F0060TEK} shows the waveform (run 60) which we will use to illustrate the analysis. It is a favorable instance of the least favorable case (highest beat rate). We took 10 runs for each of the 3 gaps (wire pairs). `Gap\,1' refers to wires 1,2 and so on. 

By way of illustration we use working graphics which the program produces during analysis for any selected run. With graphics turned off, it takes about a second to fully analyze the 30 runs.

\section{Analysis}\label{sec:analysis}
To extract frequencies from time series, the mind turns to FFT. We experi\-mented with the FFT function of the TBS1202B scope, and did observe frequency peaks decaying with time. However, that approach did not feel promising, and we have not pursued it, though by no means ruling it out in future.

We also considered nonlinear least-squares fitting with amplitudes, frequencies,
phases and decay constants as adjustable parameters. We have not ruled that out either, but to have any chance of working it would need good starting values. That consideration led us to a straightforward (if rather lengthy) analysis in the time domain, which seems sufficient on its own, and which we now describe.

Two different frequencies $f_\mathrm{high}$ and $f_\mathrm{low}$ combined linearly yield oscillations at a frequency $f_\mathrm{base}$ (there does not seem to be a standard term for this) modulated at a frequency $f_\mathrm{beat}$, and
\begin{eqnarray}\label{eqn:beats}
f_\mathrm{high}&=\;f_\mathrm{base}+f_\mathrm{beat}\\
f_\mathrm{low}&=\;f_\mathrm{base}-f_\mathrm{beat}
\end{eqnarray}
Thus we need to find the base and beat frequencies from the stored waveform.

\subsection{Base Frequency}
\begin{figure}[p]
\centering\includegraphics[height=3.5in]{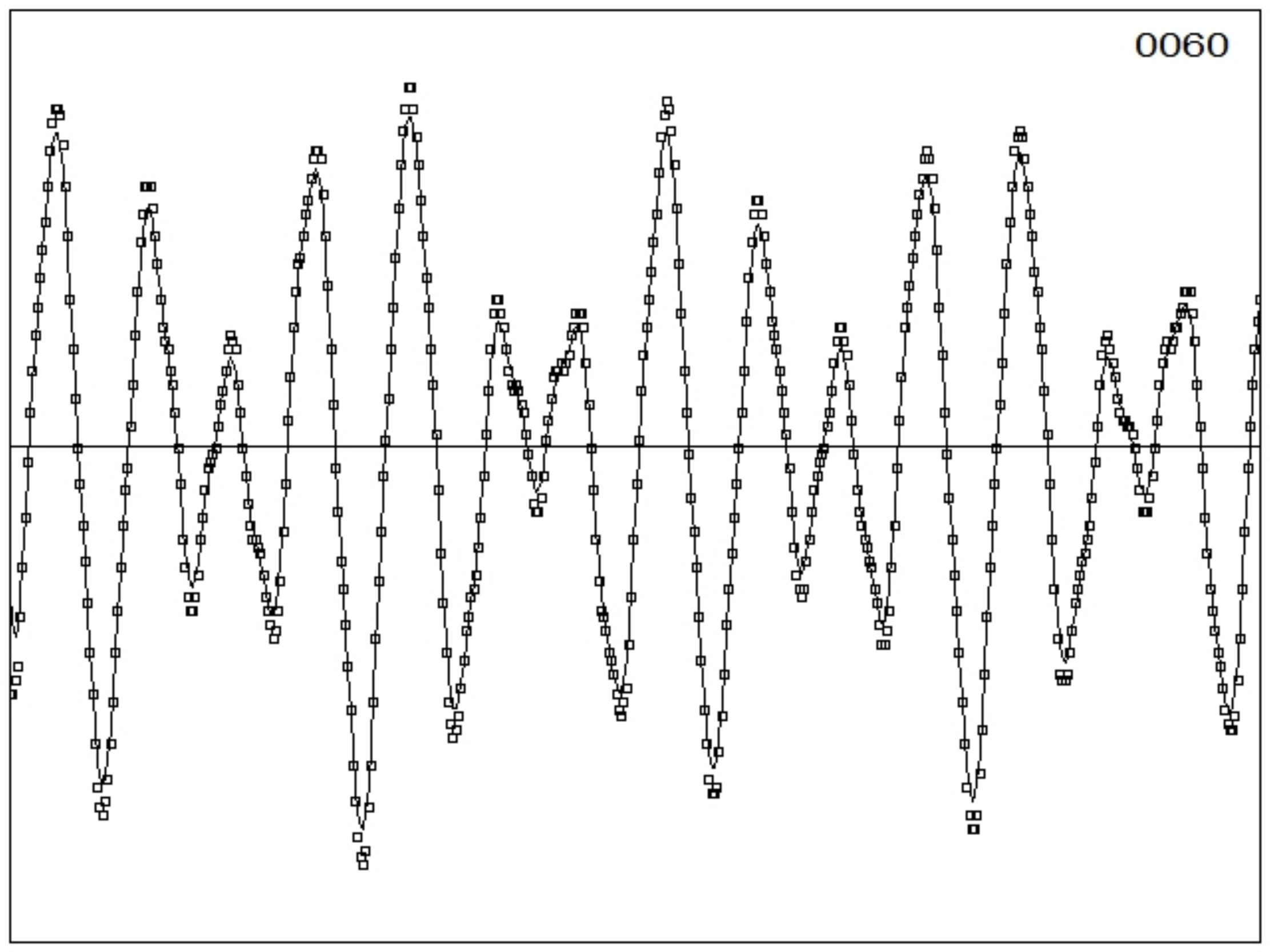}
\caption{Zoomed-in raw (points) and smoothed data (line) obtained by convolution with a Gaussian whose $\sigma$ equals 2 sampling intervals.}\label{fig:ShowTwo2}
\end{figure}

\fig{ShowTwo2} zooms in to a few beat cycles. The points represent the voltage signal $v_i$ with any DC component removed by finding the mean and subtracting it. Next we smooth (filter) the signal by convolution with a tabulated 1D normal Gaussian to reduce spurious zero crossings:
\begin{equation}
v'_i\,=\,\sum_{j\,=\,i-5\sigma}^{i+5\sigma}v_j\;G(i-j;\sigma)
\end{equation}
 $\sigma$ (a tunable parameter) is, in \fig{ShowTwo2}, equivalent to 2 sampling intervals, but it need not be integral.\footnote{~There are, in all, three tunable parameters: two smoothing time constants and a frequency distribution cut. Once optimized, these are held the same for all 30 measurements in a set.}

\begin{figure}[p]
\centering\includegraphics[height=3.5in]{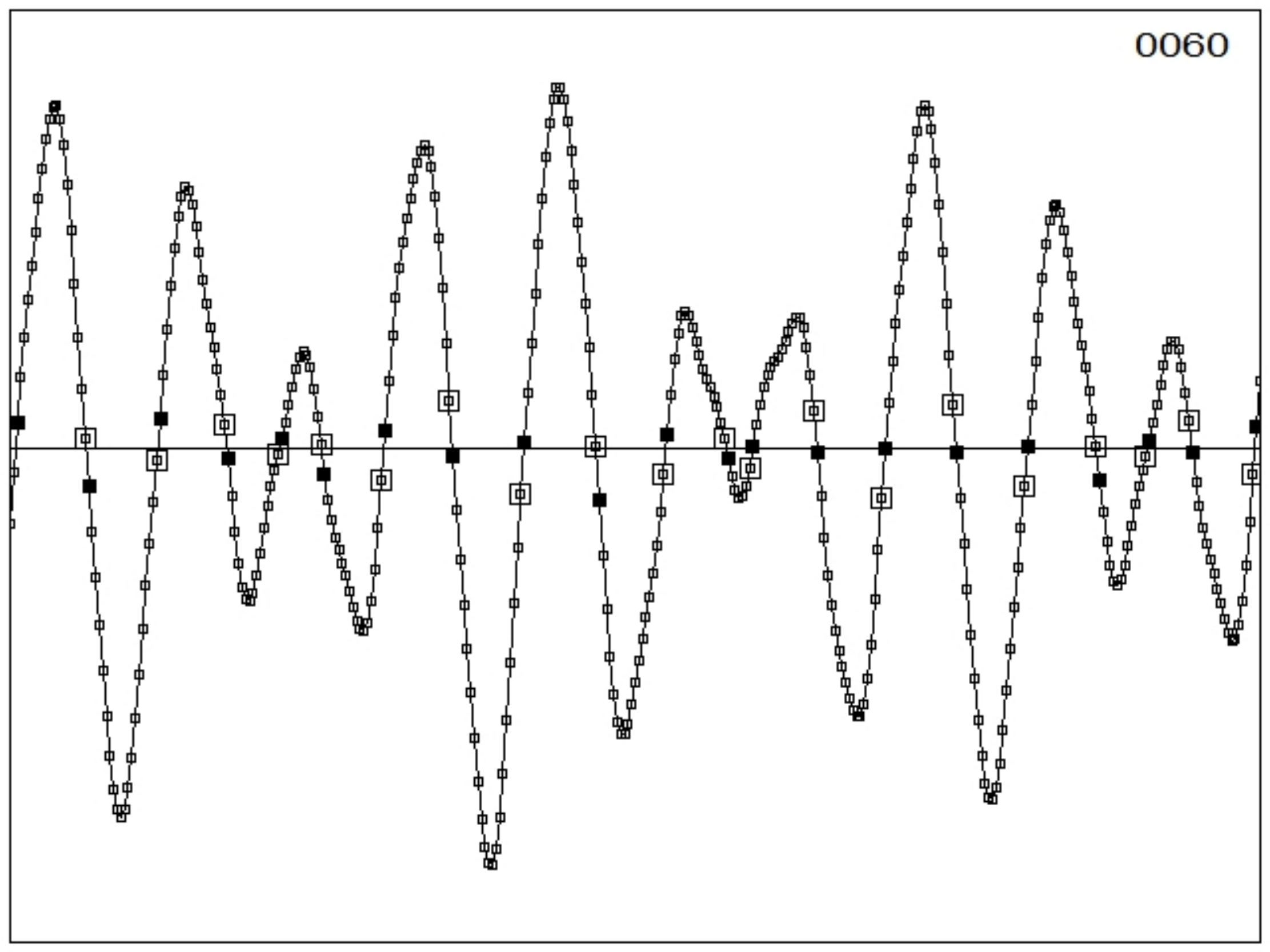}
\caption{The same data with zero crossings marked. Full, open boxes are first, last points within a half-period (HP). A width $w$ (distance between zero crossings), $x$ value (mean of interpolated zero crossings) and $y$ value ($y_{rms}$ of all points contained) are assigned to each HP.}\label{fig:ShowHP1}
\end{figure}

\fig{ShowHP1} shows the zero-crosser at work. It divides the waveform into half-periods (HPs), the full and hollow squares representing the first and last points within each HP. (Occasionally these may be the same point.) The `width' of each HP is defined as the distance between actual zero crossings obtained by linear interpolation.

\begin{figure}[p]
\centering\includegraphics[height=3.5in]{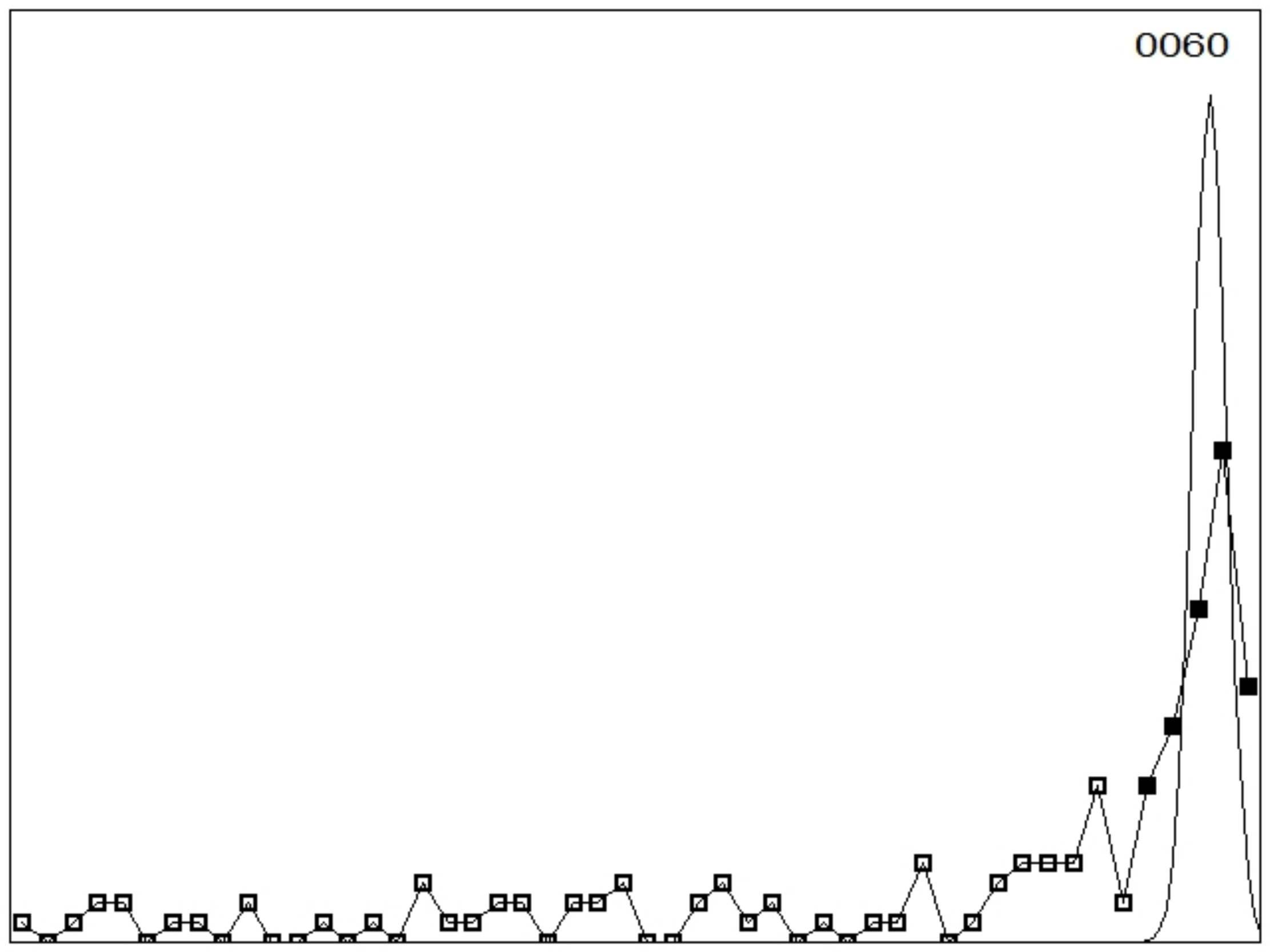}
\caption{Frequency distribution of zero crossings. Many spurious small values come from beat minima. Full boxes mark channels included in the two-step computation of the mean. An equivalent Gaussian (same area, mean and $\sigma$) is shown.}\label{fig:FreqDist1}
\end{figure}

We now have a list of widths. We form their frequency distribution (\fig{FreqDist1}) and find a peak corresponding to $f_\mathrm{base}$. When the beat frequency is high, as here, there will occur spuriously small HP widths arising from beat minima (cf. \fig{ShowHP1}). To ignore these we find the highest peak, compute its mean over channels (peak channel)$\,\pm\,$NINT(cut), then recompute it over (mean)\,$\pm$\,(cut) using split channels {\em pro rata}..\footnote{~NINT\,$\equiv$\,nearest integer.} The cut is a tunable parameter. The peak is far more prominent, and the cut less critical, at lower beat rates  because there are more $f_\mathrm{base}$ cycles between beat minima.

\subsection{Beat Frequency}
\begin{figure}[p]
\centering\includegraphics[height=3.5in]{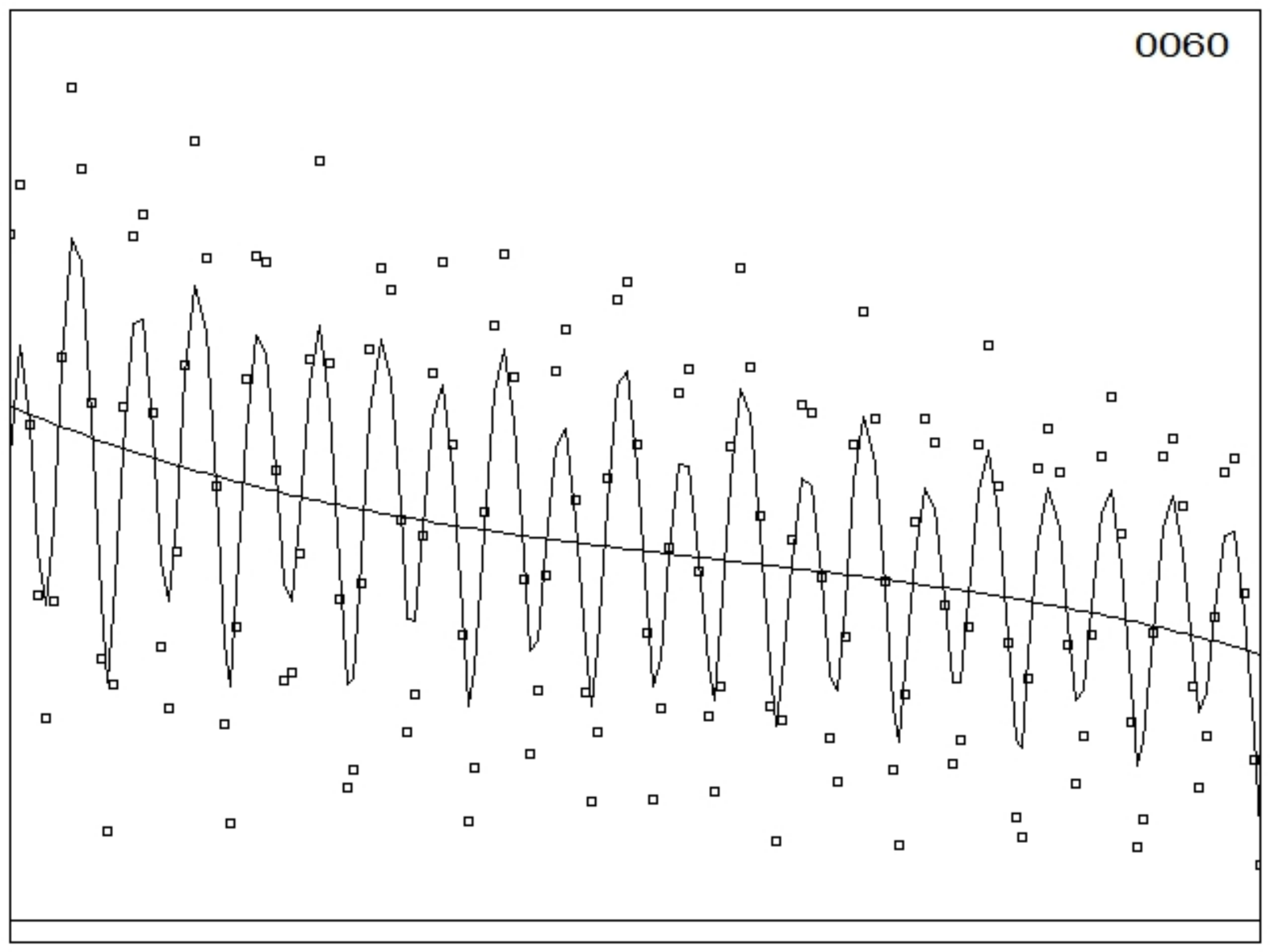}
\caption{{\em rms} values for each base half-period (points) as smoothed (line) and fit by a  cubic (four-term) polynomial.}\label{fig:ShowTwo3}
\end{figure}

\begin{figure}[p]
\centering\includegraphics[height=3.5in]{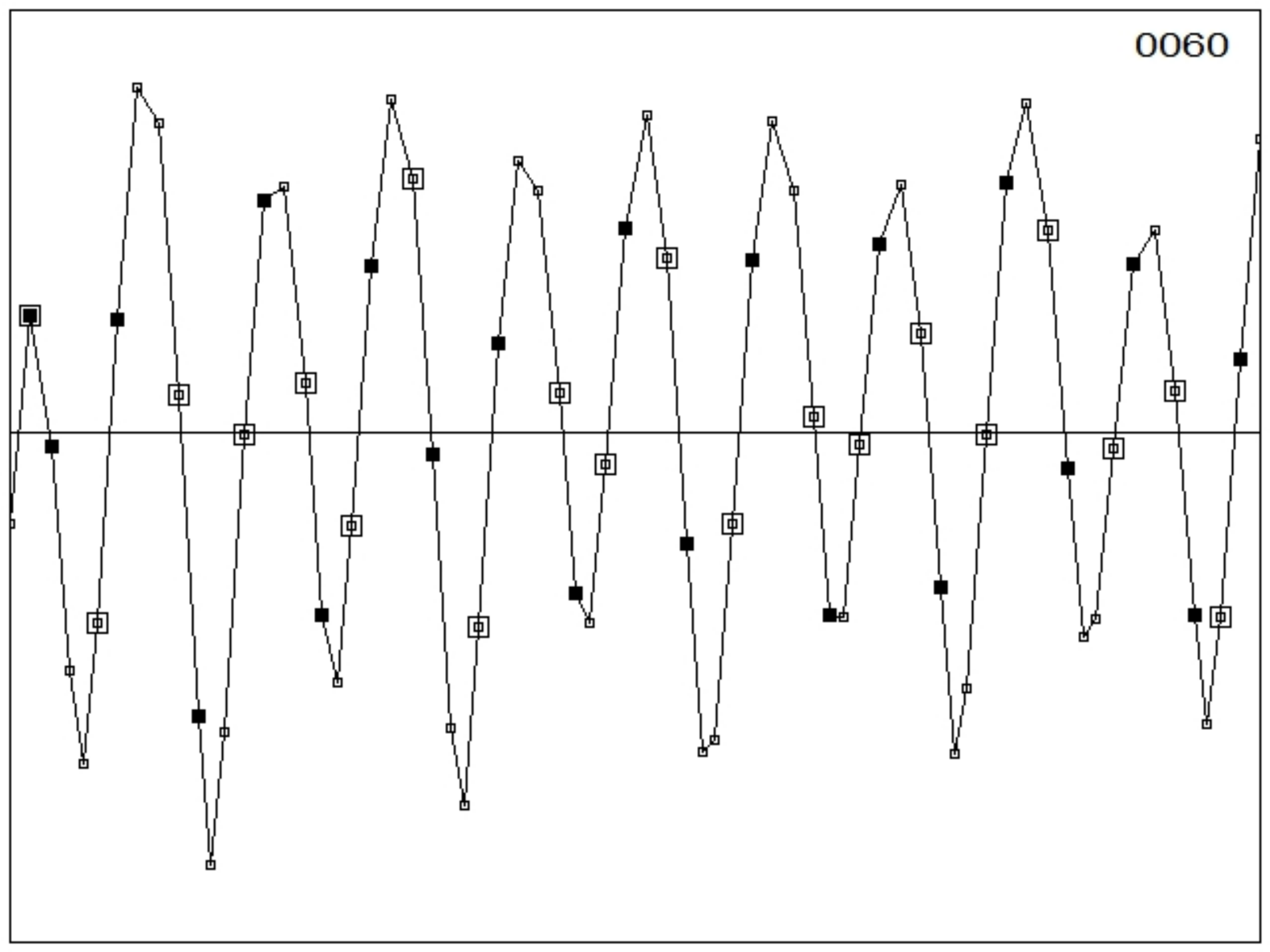}
\caption{Zero crossings corresponding to the beat `half-period'. Full, empty squares mark the first and last points, which may coincide. Note odd-even effect in widths, already evident in \fig{ShowTwo3}.}\label{fig:ShowHP2}
\end{figure}

\begin{figure}[p]
\centering\includegraphics[height=3.5in]{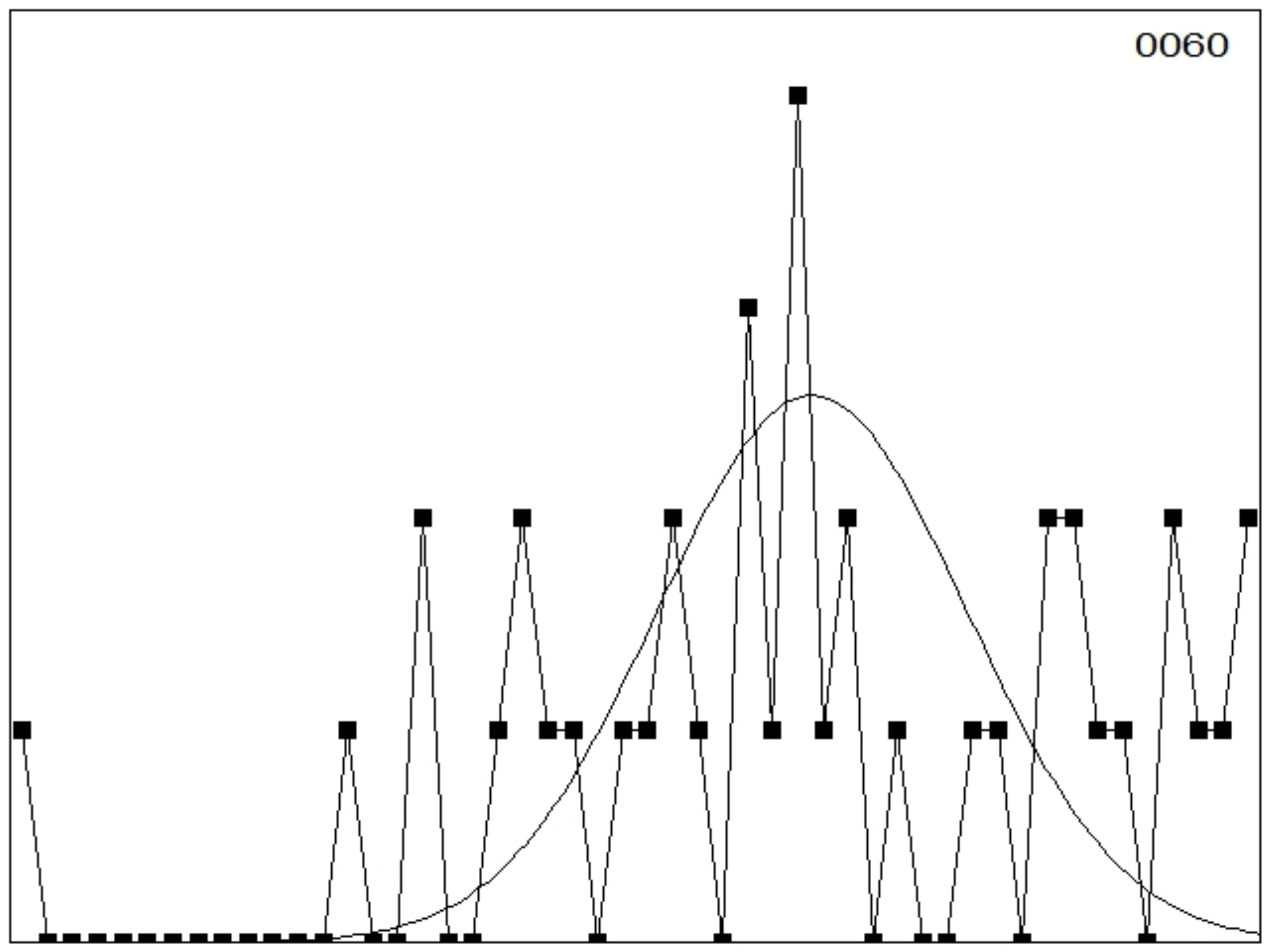}
\caption{Frequency distribution of beat half-period widths. In this case the mean should be unrestricted, and the `half-period' is actually a quarter-period.}\label{fig:FreqDist2}
\end{figure}

The most involved part of the computation is finding the waveform envelope and $f_\mathrm{beat}$ using all available points for better statistics.\footnote{~Simply finding extrema does not work well because it relies on too few points.} We compute $y_{rms}$ over all points of each HP (cf. \fig{ShowHP1}) and assign that value to the mean $x$ of that HP. That yields a new, greatly reduced, table which (after smoothing again) we fit with a cubic (four term) polynomial to account for damping (\fig{ShowTwo3}). We then find zero crossings of the fit residual, corresponding to the HPs of $f_\mathrm{beat}$ (\fig{ShowHP2}). Once again we form a frequency distribution (\fig{FreqDist2}), but we now take the unrestricted mean (no cuts).

Because we---in effect---rectify the original waveform to find the beat HPs, they show a systematic odd-even effect (cf. \fig{ShowHP2}) and therefore we must, in our final tally, drop one HP if necessary to obtain an even number. Also, `half-period' is a misnomer in this case since one full beat envelope (a full sine wave) comprises {\em four} such `half-periods'. We have retained the `HP' notation in our program, because the subroutine is the same.

\subsection{Final Steps}%%%
\begin{figure}[p]
\centering\includegraphics[height=3.5in]{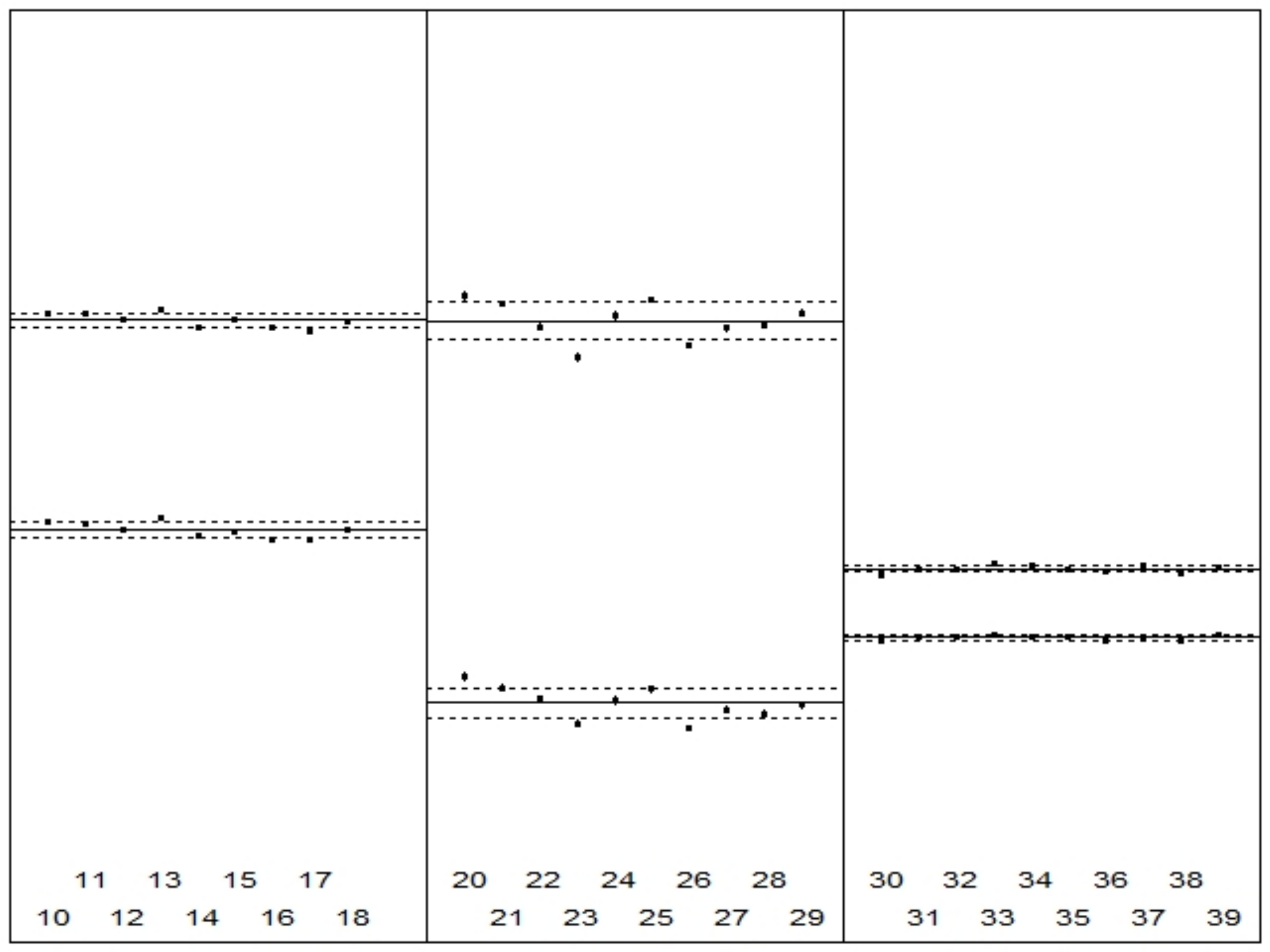}%%%
\caption{Poor outcome from the 2.5\,s series. Spacing between levels reflects $f_\mathrm{beat}$, position up/down reflects $f_\mathrm{base}$. Both are well determined for gaps 1 and 3 , but no combination of analysis parameters makes gap\,2 agree with both. Individual error bars are from the spread in the mean half-periods for each scope trace (see text). Dashed lines are the $\pm1\sigma$ spreads in the runs for each gap (listed at bottom).}\label{fig:BadFinal}
\end{figure}

\begin{figure}[p]
\centering\includegraphics[height=3.5in]{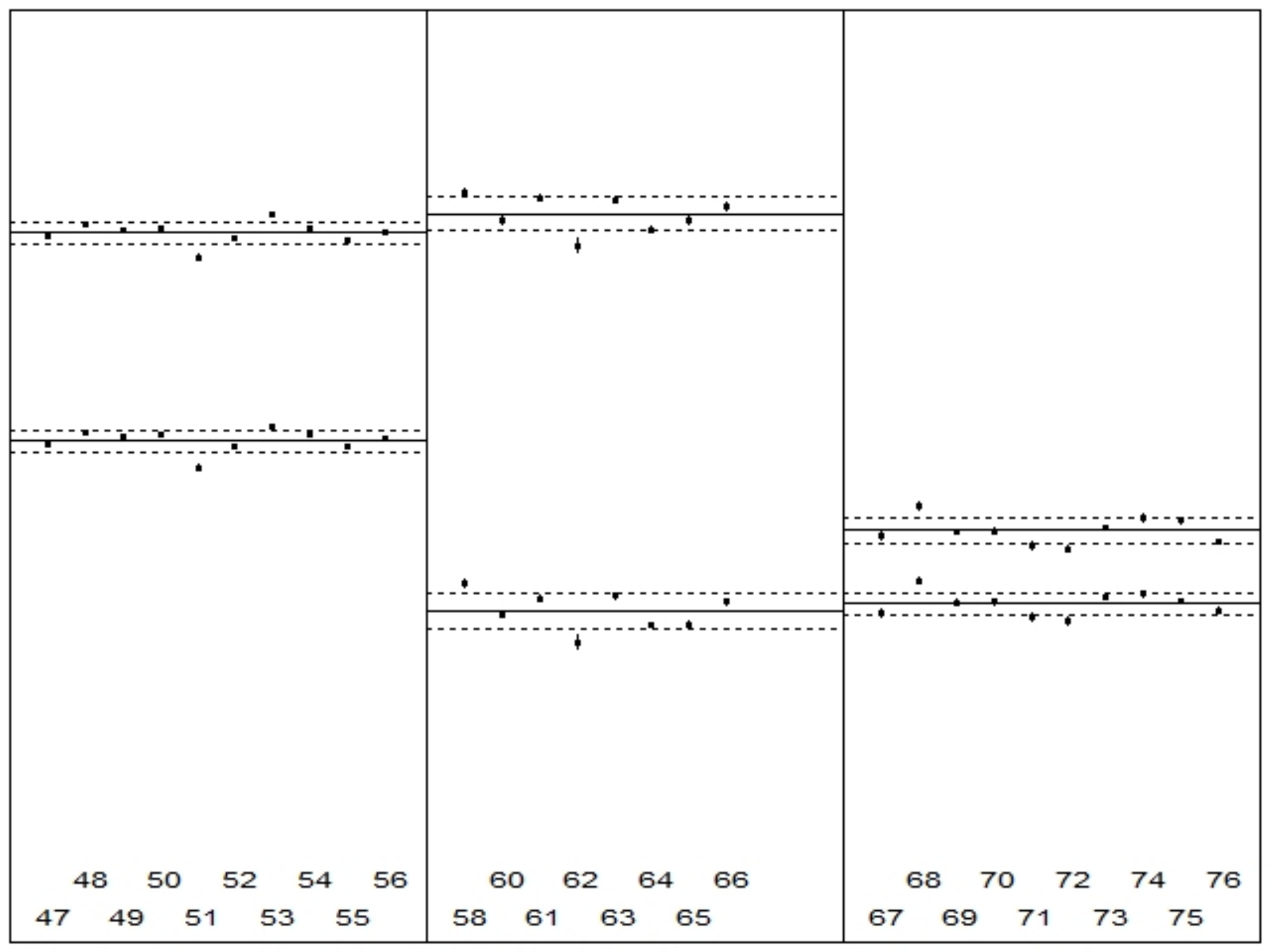}
\caption{Good outcome from the 1\,s series. It takes about 1s to compute this result from 28 measurements. Frequency zero is suppressed, as in \fig{BadFinal}.}\label{fig:ShowFinal}
\end{figure}

\begin{figure}[p]
\centering\includegraphics[height=3.5in]{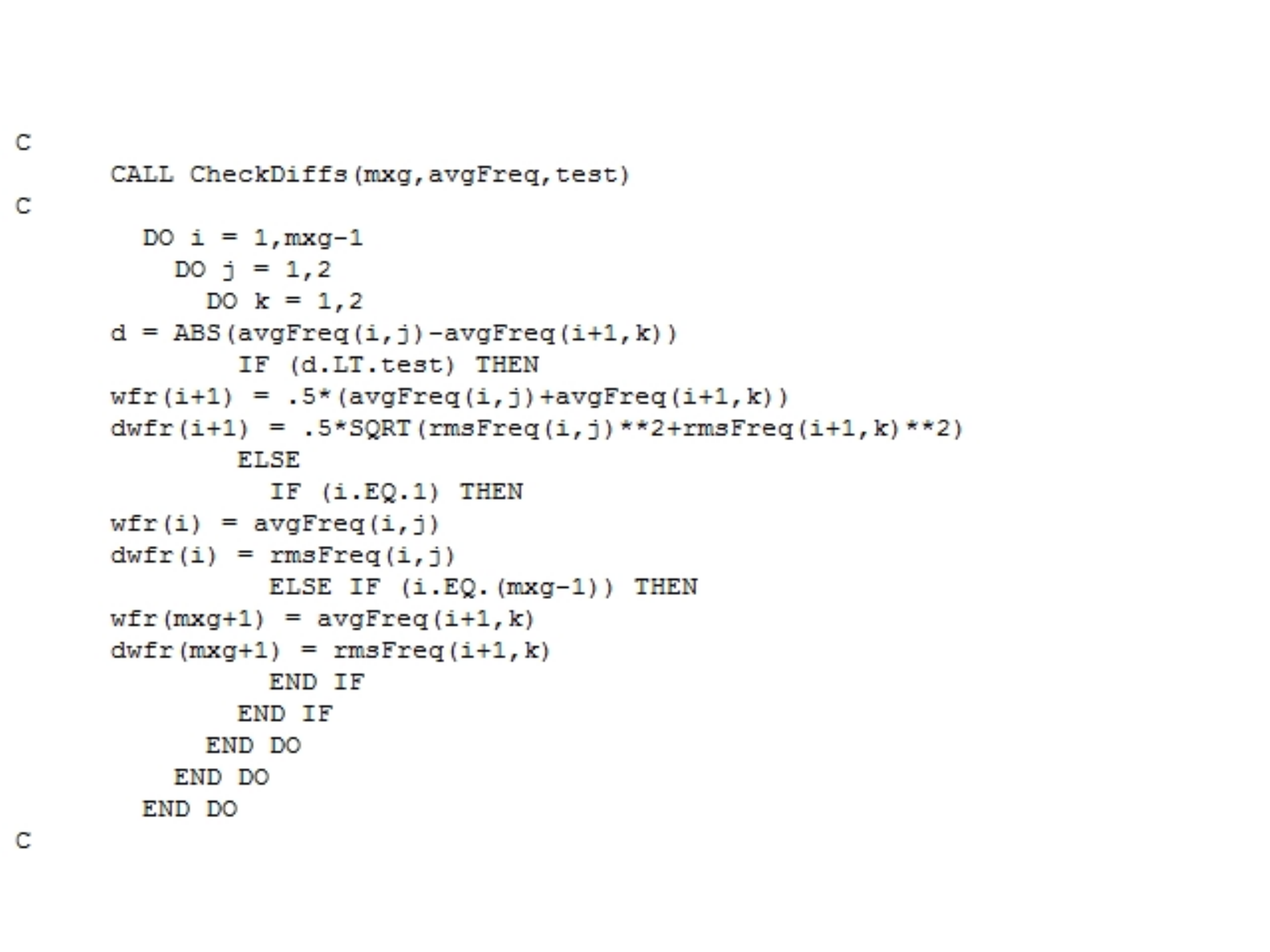}
\caption{Logic to convert (disambiguate) two $f$s by gap to one $f$ per wire.}\label{fig:Disambiguate}
\end{figure}

At this stage we know the half-period corresponding to $f_\mathrm{base}$ and the quarter-period corresponding to $f_\mathrm{beat}$. For each run, the final half- or quarter-period is the mean of $n$ entries in a frequency distribution (cf. \fig{FreqDist1} and \fig{FreqDist2}). Therefore the standard error in the mean (from each scope trace) is
\[\sigma_\mathrm{mean}\;=\;n^{-1/2}\,\sigma\]
where $\sigma$ is the {\em rms} spread of the selected region. From the periods, we find the base and beat frequencies ($f=1/T$), then $f_\mathrm{high}$ and $f_\mathrm{low}$ from \eqn{beats} and their errors by propagation of errors. We sort the results on the key $(100\times n_\mathrm{gap}+n_\mathrm{data file})$ to generate (independent of analysis order) \fig{BadFinal} (2.5\,s series, a poor result) and \fig{ShowFinal} (1\,s series, a good result). \tbl{resultsByGap} lists the 1\,s numbers by gap.
\begin{table}[h]
\setlength{\tabcolsep}{5pt}
\begin{center}\begin{tabular}{crrrr}
\multicolumn{1}{c}{gap}&           
\multicolumn{1}{c}{$f_\mathrm{high}$}&
\multicolumn{1}{c}{$1\,\sigma$}&           
\multicolumn{1}{c}{$f_\mathrm{low}$}&
\multicolumn{1}{c}{$1\,\sigma$}\\           
\noalign{\vskip5pt}          
  1&   73.042&    0.561&   61.895&    0.581\\
  2&  74.055&    0.915&   52.790&    0.975\\
  3&   57.075&    0.673&   53.157&    0.608
\end{tabular}\end{center}
\caption{1\,s results by gap (Hz).\label{tbl:resultsByGap}}
\end{table}
We then match up those six numbers, averaging two of them for `internal' wires, to obtain the frequency and its error for each wire. \fig{Disambiguate} is the corresponding program fragment. \texttt{CheckDiffs} is a procedure which examines frequency differences between adjacent gaps and returns an appropriate value as \texttt{test}. Finally, we compute the tension in each wire and its error using \eqn{f} and propagation of errors. \tbl{resultsByWire} lists the 1\,s results by wire.
\begin{table}[h]
\setlength{\tabcolsep}{5pt}
\begin{center}\begin{tabular}{crrrr}
\multicolumn{1}{c}{wire}&           
\multicolumn{1}{c}{$f$}&
\multicolumn{1}{c}{$1\,\sigma$}&           
\multicolumn{1}{c}{$T$}&
\multicolumn{1}{c}{$1\,\sigma$}\\           
\multicolumn{1}{c}{}&           
\multicolumn{1}{c}{(Hz)}&
\multicolumn{1}{c}{(Hz)}&           
\multicolumn{1}{c}{(N)}&
\multicolumn{1}{c}{(N)}\\           
\noalign{\vskip5pt}          
  1&   61.895&    0.581&    5.189&    0.097\\
  2&   73.549&    0.537&    7.327&    0.107\\
  3&   52.973&    0.575&    3.801&    0.082\\
  4&   57.075&    0.673&    4.412&    0.104
\end{tabular}\end{center}
\caption{1\,s results by wire.\label{tbl:resultsByWire}}
\end{table}

\section{Direct Measurement of Tension}\label{sec:direct}
\begin{figure}[p]
\centering\includegraphics[height=3.5in]{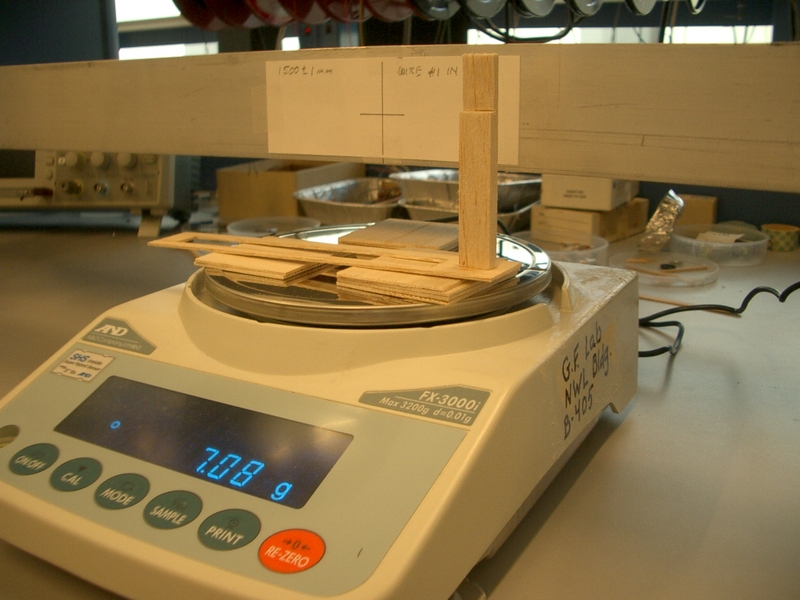}
\caption{Balsa strut and holder posed on scale pan with shims.}\label{fig:strut}
\end{figure}

\begin{figure}[p]
\centering\includegraphics[height=3.5in]{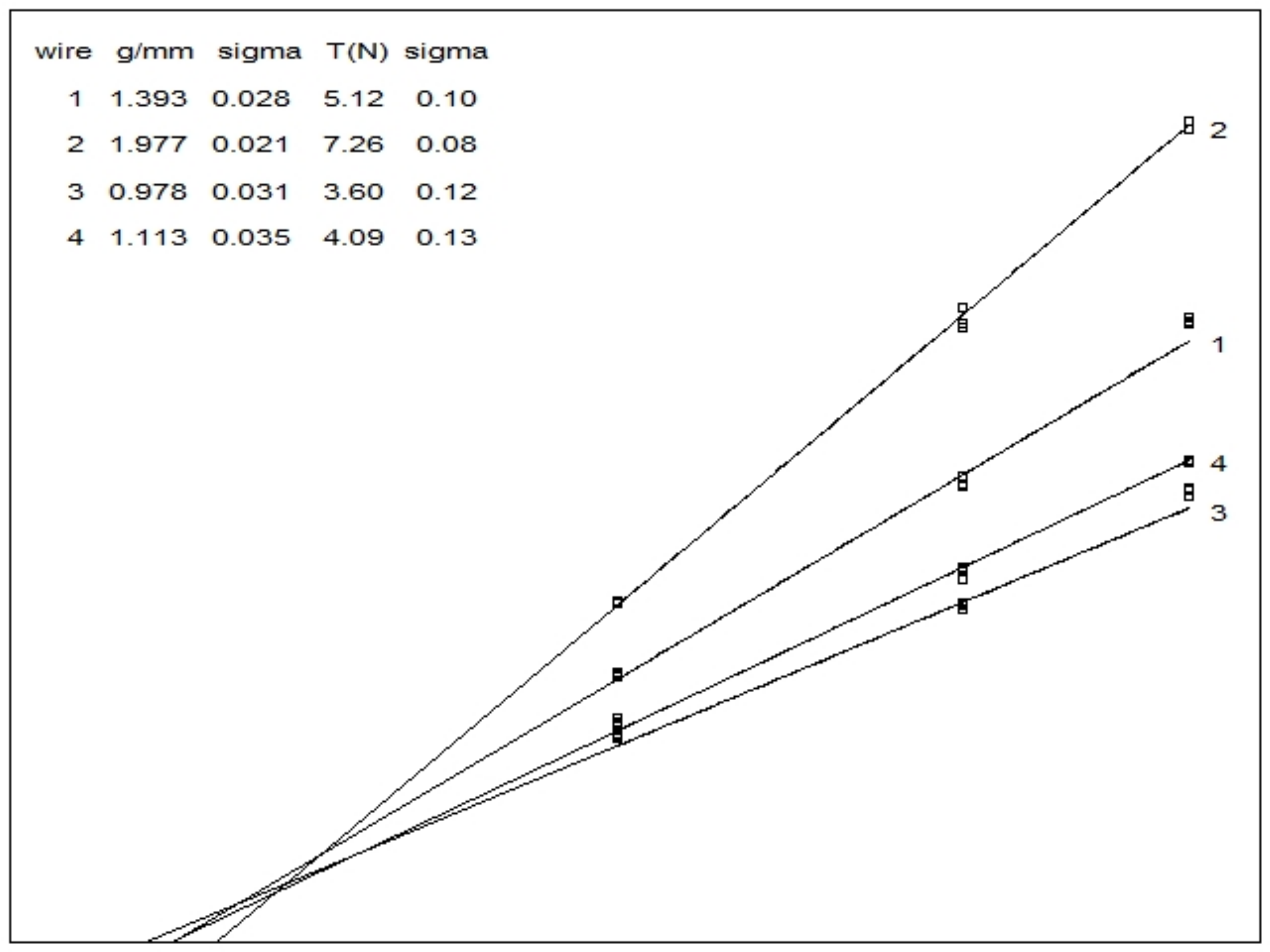}
\caption{g/mm data with fits and results.}\label{fig:gpmm}
\end{figure}

\begin{figure}[p]
\centering\includegraphics[height=4.5in]{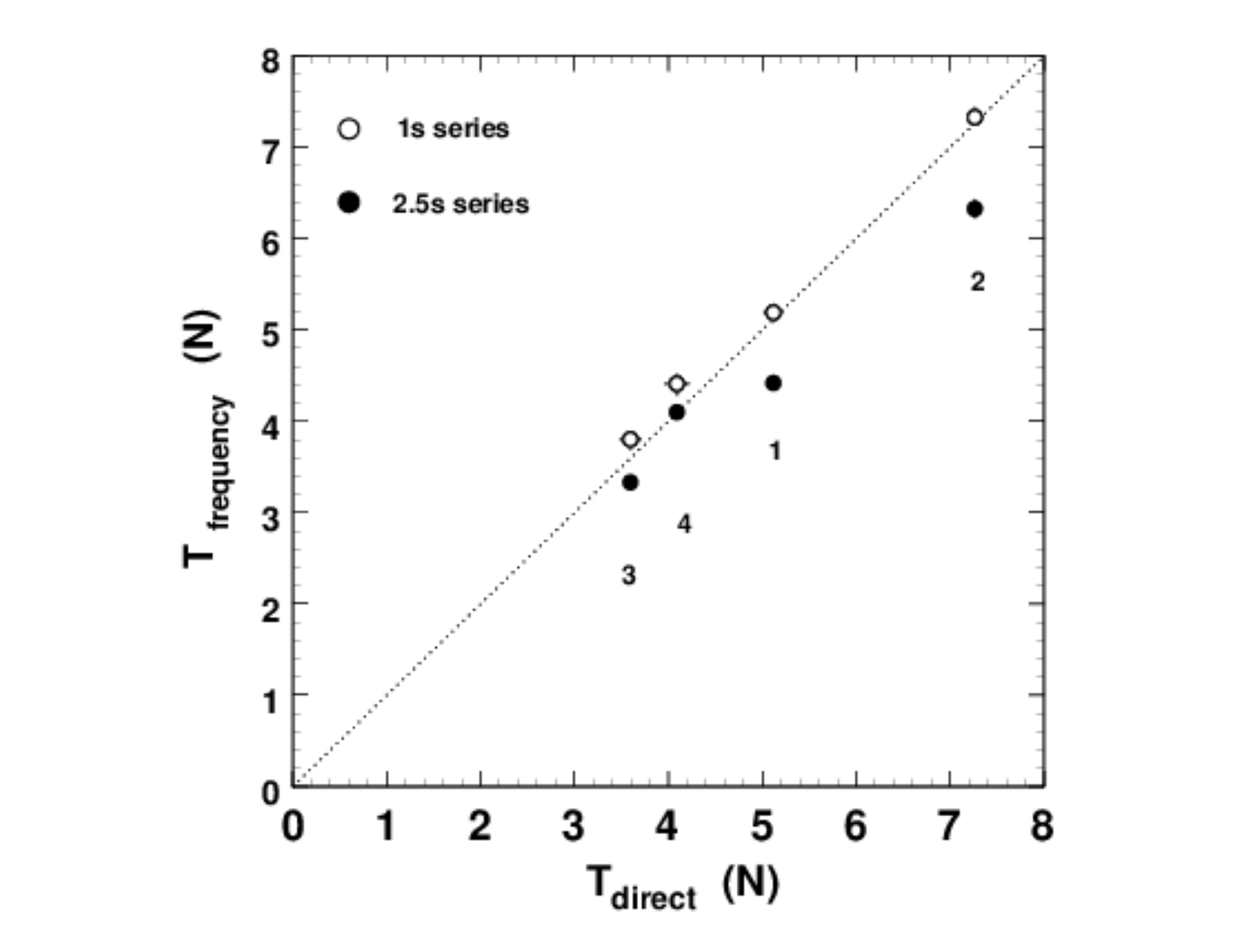}
\caption{Grand summary: plot of Table \ref{tbl:GrandSummary}. Tension from the direct measurement is taken as ground truth and tension derived from measured resonant frequency is plotted against it. Dashed line is the ideal relation.}\label{fig:GrandSummary}
\end{figure}

To measure the tension directly, we mounted the jig upside-down over a laboratory balance and used a combination of three shims and a strut (all balsa-wood) to measure the (g/mm) required to deflect each wire at its midpoint (\fig{strut}). We took three sets of five readings per wire. \fig{gpmm} shows the fits\footnote{~A quadratic fit shows some curvature, which we do not understand and have ignored here. The non-zero intercept merely reflects the strut travel before it encounters the wire. The strut was designed to bottom out and protect the wire if too much travel was attempted.} and the final results, from
\[T\;=\;9.8\times\frac{L}{4}\times\hbox{(g/mm)}\qquad\hbox{N}\]
with $L$ in m. This measurement was mainly for fun; nobody seriously doubts \eqn{f}. \tbl{GrandSummary} and \fig{GrandSummary} summarize our four-wire study. The frequency based and direct tensions agree well for the 1\,s series and poorly for the 2.5\,s series.
\begin{table}[h]
\setlength{\tabcolsep}{5pt}
\begin{center}\begin{tabular}{crrrrrr}
\multicolumn{1}{c}{wire}&           
\multicolumn{1}{c}{$T_\mathrm{direct}$}&
\multicolumn{1}{c}{$1\,\sigma$}&           
\multicolumn{1}{c}{$T_\mathrm{1\,s}$}&
\multicolumn{1}{c}{$1\,\sigma$}&           
\multicolumn{1}{c}{$T_\mathrm{2.5\,s}$}&
\multicolumn{1}{c}{$1\,\sigma$}\\ 
\noalign{\vskip5pt}          
    1&    5.120&    0.102&    5.189&    0.097&    4.418&    0.062\\
    2&    7.264&    0.077&    7.327&    0.107&    6.326&    0.100\\
    3&    3.596&    0.115&    3.801&    0.082&   3.330&    0.056\\
    4&    4.092&    0.129&    4.412&    0.104&    4.099&    0.025
\end{tabular}\end{center}
\caption{Grand summary by wire: tensions and 1$\sigma$ errors (N) from direct measurement, 1\,s scope measurements and 2.5\,s scope measurements.\label{tbl:GrandSummary}}
\end{table}

\section{Summary and Discussion}\label{sec:discussion}
We have described a method which, with some obvious improvements, would seem to be a viable alternative to frequency search techniques. Wires are set in motion by tapping the frame, and a `capacitance microphone' signal is recorded using a preamp and digital oscilloscope. Ambient EMI is rejected by designing the preamp to respond only to the differential signal from a wire pair. If the tensions in the wires are different, there will be beats in the signal. A straightforward analysis extracts the individual resonant frequencies and, from them, the tensions. In our demonstration, four tensions agreed well with direct measurement. 

If the oscilloscope or other data acquisition device were programmed as an extension of the computer, one could measure wire tension to roughly $\pm2\%$ in about 1\,s per wire with simple equipment, off-the-shelf except for the preamp.

In the system envisioned for DUNE \cite{Garcia-Gamez2019}, the comparative slowness of the frequency-search method is overcome by examining many wires in parallel. Of course, two can play at that game, but the `bespoke' character of the equipment escalates rather rapidly: multiple contacts to multiple pads or solder bumps, multiple high voltage switches etc. The usefulness (if any) of the method described herein may consist in getting a very rapid evaluation of a few wires with very simple equipment. 

\section{Acknowledgements}
We thank Harvard, the Physics Department and the Laboratory for Particle Physics and Cosmology (LPPC) for generous and sustained support. The administrative staff of the Department of Molecular and Cellular Biology, where our office happens to be, has been unfailingly kind. We also thank Sebastien Prince for the wire, John Oliver for reminding us of EMI, Scott Fulton for the loan of his laboratory balance, Steve Sansone for contributions of stock, and Ethan Cascio for introducing us to KiCad.

%\bibliographystyle{unsrt}\bibliography{/pctexv4/work/pbs/master}

\begin{thebibliography}{1}

\bibitem{Garcia-Gamez2019}
D. Garcia-Gamez, V. Basque, T.G. Brooks, J.J. Evans, M. Perry, S.
  S{\"o}ldner-Rembold, F. Spagliardi and A.M. Szelc, `A novel electrical method
  to measure wire tensions for time projection chambers,' Nucl. Instr. Meth.
  Phys. Res. {\bf A915} (2019) 75-81,
  https://doi.org/10.1016/j.nima.2018.09.031.

\bibitem{Lang1999}
K. Lang, J. Ting and V. Vassilakopoulos, `A technique of direct tension
  measurement of a strung fine wire,' Nucl. Instr. Meth. Phys. Res. {\bf A420}
  (1999) 392-401.

\bibitem{Brinkley1996}
B. Brinkley, J. Busenitz, G. Zilizi, `Wire tension measurement using voltage
  switching,' Nucl. Instr. Meth. Phys. Res. {\bf A373} (1996) 23-29.

\bibitem{Gould2005}
H. Gould, J. Tobochnik and W. Christian, `Simulating particle motion,'
  evidently Chapter\,3 of ``Introduction to Computer Simulation Methods in
  Physics'', now out of print.

\bibitem{Cromer1981}
Alan Cromer, `Stable solutions using the Euler approximation,' American Journal
  of Physics {\bf49} (1981) 455-459; doi: 10.1119/1.12478.

\bibitem{Purcell}
Edward M. Purcell, `Electricity and Magnetism,' $2^\mathrm{nd}$ Ed.,
  McGraw-Hill (1985).

\end{thebibliography}

%\clearpage\listoffigures

\end{document}